\documentclass[preprint,12pt]{elsarticle}
\biboptions{sort&compress}
\usepackage{url}
\usepackage{subfig}
\usepackage{amsfonts}
\usepackage{amssymb}
\usepackage{amsmath}
\usepackage{threeparttable}
\usepackage{multirow}
\usepackage{lineno}
\usepackage[colorlinks]{hyperref}
\usepackage{listings}
\usepackage[scale=0.8]{geometry}
\usepackage{color}
\usepackage{mathtools} 

\journal{Computer Physics Communications}
\begin{document}

\begin{frontmatter}
		\title{varRhoTurbVOF: a new set of volume of fluid solvers for turbulent isothermal multiphase flows in OpenFOAM\tnoteref{t1}}
		\tnotetext[t1]{\textcopyright 2019. This manuscript version is made available under the CC-BY-NC-ND 4.0 license \url{http://creativecommons.org/licenses/by-nc-nd/4.0/}}
		\author[a]{Wenyuan~Fan\corref{cor1}}
		\ead{wf@kth.se}
		\author[a,b]{Henryk~Anglart}
		\cortext[cor1]{Corresponding author.}
		\address[a]{Nuclear Engineering Division, Department of Physics, KTH Royal Institute of Technology, 106 91 Stockholm, Sweden}
		\address[b]{Institute of Heat Engineering, Warsaw University of Technology, 21/25 Nowowiejska Street, 00-665 Warsaw, Poland}
		
		\begin{abstract}
			The volume of fluid (VOF) method is a popular approach for multiphase flow modeling. The open-source computational fluid dynamics (CFD) software, OpenFOAM, implements a variety of VOF-based solvers and provides users a wide range of turbulence models. Since isothermal multiphase flows under the VOF framework belong to the variable-density incompressible flow category, the isothermal VOF-based solvers in OpenFOAM fail to use the correct turbulence models. varRhoTurbVOF is designed to solve this issue and with the hope to replace all the corresponding existing solvers in the future. With the object-oriented paradigm, varRhoTurbVOF guarantees the usability, reusability and maintainability of the codes. Aside from turbulence modeling, all other features in the original solvers are preserved in varRhoTurbVOF.
		\end{abstract}
		
	\begin{keyword}
		VOF \sep CFD \sep turbulence modeling \sep variable-density incompressible flow \sep OpenFOAM
	\end{keyword}
	
\end{frontmatter}


{\bf PROGRAM SUMMARY}

\begin{small}
	\noindent
	{\em Program Title: varRhoTurbVOF} \\
	{\em Licensing provisions: GPLv3}\\
	{\em Programming language: C++}\\	
	{\em Supplementary material: http://dx.doi.org/10.17632/7mp25kyb4p.4}\\	
	{\em Nature of problem:}\\
	Under the VOF framework, the flow of the isothermal mixture belongs to the variable-density incompressible flow category. For such flows, VOF-based solvers of OpenFOAM fail to construct the correct governing equations for turbulence modeling. varRhoTurbVOF contains a set of newly designed VOF-based solvers which could use the desired governing equations for turbulence quantities.\\
	{\em Solution method:}\\
	varRhoTurbVOF creates a new class for variable-density incompressible turbulence models, which allows reusing the existing turbulence model template classes. A set of VOF-based solvers are then created to be able to construct variable-density incompressible turbulence models.
\end{small}
\sloppy
\section{Introduction}
\label{sect:introduction}
Multiphase flows, e.g. air-water flows in oceans and gas-oil flows in oil transfer lines, are often encountered in nature and in industrial applications. However, multiphase flow modeling is challenging due to the existence of the moving interface between different phases. Among various modeling approaches, the volume of fluid (VOF) method \citep{Hirt1981} is a popular and widely adopted one due to the reasons that will be discussed in Section \ref{sect:vof}. As a matter of fact, many open-source multiphase computational fluid dynamics (CFD) codes, e.g. OpenFOAM \citep{Weller1998,Ubbink1999,Jasak2009} and Gerris \citep{Popinet2003}, have implemented solvers based on the VOF method.

Even though the VOF method has significantly evolved and been extensively studied since its inception, it still has drawbacks which are mostly caused by the existence of the transition zone from one phase to another. This transition zone introduces difficulties in getting a sharp interface and calculating the interface curvature accurately. Therefore, VOF-related investigations mainly focus on sharpening the interface and calculating the curvature accurately \citep{Kothe1998,Tucker2003,Cummins2005,Ito2013,Roenby2016,Scheufler2019}. In such studies, the flow field is either negligible or known as \textit{a priori}. The reason is that there are available analytical solutions, e.g. Young-Laplace equation and the solution provided by \citep{Prosperetti1981}, or well defined benchmarks for such flow conditions. These simple test cases play an significant role in identifying issues with existing algorithms, validating newly designed techniques and advancing the VOF method.

The stable progress in the VOF method unfortunately lags far behind the demands from practical applications, where the flow is often turbulent. Without dedicated turbulence models for multiphase flows, the common practice of modeling turbulent multiphase flows is combining the VOF method with single-phase turbulence models \citep{ANSYSInc.,OpenCFD2018} due to its simple formulation. Therefore, the implementation of this treatment should be a relatively easy task. However, the reality is the opposite, at least for the open-source community. For instance, no explicit turbulence models are provided in Gerris, and OpenFOAM has been using a mathematically inconsistent modeling methodology for turbulent multiphase flows under the VOF framework, as will be discussed in Section \ref{sect:issues}. Regarding the incorrect implementation in OpenFOAM, the most important reason is that there is no analytical solution for turbulent multiphase flows, which is not surprising because there is no such solution even for single-phase turbulent flows. Consequently, it is difficult to detect flaws of existing codes and verify newly designed codes. 
	
Considering the huge demand for turbulent multiphase simulations and the popularity of both OpenFOAM and the VOF method, this work aims at providing a better open-source OpenFOAM- and VOF-based platform to the multiphase modeling community. The paper is structured as follows: Section \ref{sect:vof} and Section \ref{sect:overview} briefly introduce the physics behind the VOF method, turbulence modeling, and their combination in OpenFOAM; Section \ref{sect:issues} points out the limitations of current isothermal VOF-based solvers in OpenFOAM; Section \ref{sect:code} provides the philosophy, implementation and verification of the newly designed solvers; Section \ref{sect:usage} describes the usage of new solvers; Section \ref{sect:performance} conducts a performance evaluation for the new and old solvers; Section \ref{sect:conclusion} summarizes the work and provides outlooks.

\section{Overview of the VOF method}
\label{sect:vof}

The VOF method \citep{Hirt1981} was first developed to model immiscible two-phase flows with a simple concept for interface advection:
\begin{equation}
\label{eq:VOF}
\frac{\partial \alpha}{\partial t}+\vec{u} \cdot \nabla  \alpha=0,
\end{equation} 
where $\alpha$ is the volumetric fraction of the primary phase in a control volume (cell). $\alpha=1$ means that the cell is entirely occupied by the primary phase, and $\alpha=0$ implies that the cell is purely filled by the secondary phase. Eq. \eqref{eq:VOF} also indicates that mass conservation is always guaranteed for the two-phase system, which makes it favorable for two-phase flow simulations.  However, it is an additional treatment that makes VOF so popular. By substituting the density and viscosity in the single-phase governing equations with the mixture density:
\begin{equation}
\label{eq:average density}
\rho_m = \alpha \rho _1+(1-\alpha)\rho _2,
\end{equation}
and the mixture viscosity:
\begin{equation}
\label{eq:average viscosity}
\mu_m = \alpha \mu _1+(1-\alpha)\mu _2,
\end{equation}
where subscripts 1 and 2 denote the primary and secondary phase respectively, the resultant governing equations could be used to describe the two-phase system.

In order to illustrate an important characteristic of the VOF method, a fundamental definition in fluid mechanics, i.e. incompressible flow, is firstly introduced. A flow is incompressible if it satisfies:
\begin{equation}
\label{eq:incompressible_u}
\nabla \cdot \vec{u}  = 0,
\end{equation}
or the equivalent form:
\begin{equation}
\label{eq:incompressible_rho}
\frac{\partial \rho}{\partial t}+\vec{u} \cdot \nabla  \rho=0.
\end{equation}
Therefore, a flow with constant density is always incompressible since Eq. \eqref{eq:incompressible_rho} is automatically satisfied. This type of flow is referred to as strict incompressible flow. However, for a flow with variable density, as long as Eq. \eqref{eq:incompressible_u} is fulfilled, the flow is still incompressible, and this type of flow is referred to as variable-density incompressible flow. One reason for defining this group of flows separately is that many governing equations could be simplified by using Eq. \eqref{eq:incompressible_u}. Consequently, the computing overhead could be reduced.

For an isothermal immiscible two-phase flow system, the properties of each phase are usually assumed to be constant. Therefore the flow of an individual phase is incompressible. However, by introducing the mixture-property concept, the mixture property is changing with $\alpha$. Therefore, such two-phase flows belong to the variable-density incompressible flow category. This is an important concept which finally causes the issues with current VOF-based solvers in OpenFOAM. 

\section{Overview of turbulence modeling in OpenFOAM}
\label{sect:overview}

The turbulence modeling capability is an undeniable outstanding feature of OpenFOAM. Both the Reynolds-Averaged Navier-Stokes (RANS) approach and Large Eddy Simulation (LES) are available for turbulence modeling. Plus, hybrid approaches, e.g. Detached Eddy Simulation (DES), Delayed Detached Eddy Simulation (DDES) and Improved Delayed Detached Eddy Simulation (IDDES), could also be used for turbulence modeling. Despite the diversity of such modeling approaches, the momentum equation could always be written as
\begin{equation}
	\label{eq:momentum}
	\frac{\partial \rho \vec{u}}{\partial t}+ \nabla \cdot (\rho \vec{u} \vec{u})= -\nabla p^* + \nabla \cdot \left[ (\mu + \mu_t) \left(\nabla \vec{u}+(\nabla \vec{u})^T - \frac{2}{3} (\nabla \cdot \vec{u}) I \right) \right] + \vec{F_b},
\end{equation} 
where $I$ is the unit second-order tensor; $\vec{F_b}$ includes the gravitational force and other forces, if any; $\vec{u}$ is the time-averaged velocity for RANS and space-filtered velocity for LES; $\mu_t$ is the turbulent viscosity for RANS and subgrid-scale viscosity for LES;  $p^* = p + \frac{1}{3}\textrm{tr}(\tau_t)$ is the modified pressure with $p$ being the real pressure and $\tau_t$ being the modeled turbulent stress. 

\subsection{Turbulence models}

\label{sect:RANS}

Eq. \eqref{eq:momentum} is incomplete due to the occurrence of $\mu_t$. In order to make it complete, various turbulence models use different additional equation(s) to calculate $\mu_t$. Among various equations of different turbulent models, we consider the most representative $k$ equation in RANS modeling:
\begin{equation}
\label{eq:kEquation_full}
\frac{\partial \rho k}{\partial t}+ \nabla \cdot (\rho \vec{u} k)= \rho P - \rho \epsilon +\nabla \cdot \left[ \left(\mu + \frac{\mu_t}{\sigma_k} \right) \nabla k \right],
\end{equation} 
where $\rho P$ is the production term, $\rho \epsilon$ is the dissipation term, and $\nabla \cdot \left[ \left(\mu + \frac{\mu_t}{\sigma_k}\right) \nabla k \right]$ is the diffusion term with $\sigma_k$ being the turbulent Schmidt number for $k$.

\begin{figure}[h!]
	\centering
	\includegraphics[width=1\linewidth]{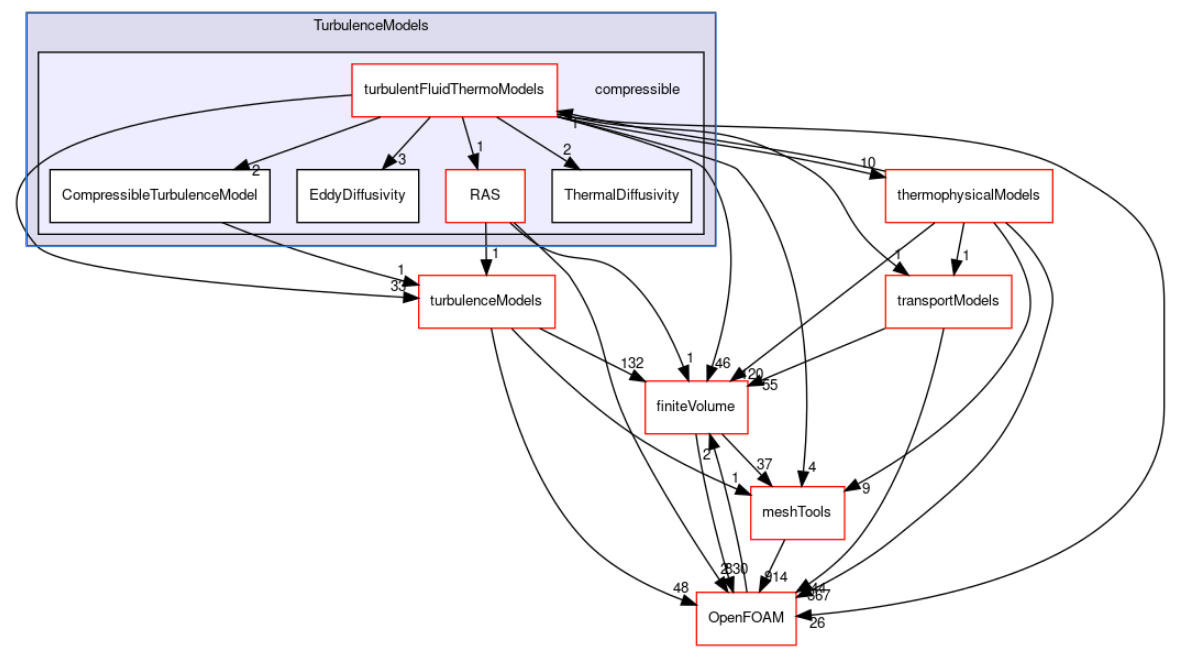}
	\caption{Directory dependency graph for compressible turbulence models in OpenFOAM v1706 \citep{OpenCFD2018}. Thermal properties are always needed to construct a compressible turbulence model.}	
	\label{fig:compressibleDependency}
\end{figure}

We refer to Eq. \eqref{eq:kEquation_full} as the full form of $k$ equation in the sense that the divergence-free condition is not used. Therefore, Eq. \eqref{eq:kEquation_full} is applicable to both compressible and incompressible flows. However, as shown in Fig. \ref{fig:compressibleDependency}, the compressible version of turbulence models is constructed as \texttt{turbulentFluidThermoModels} which means that the thermal properties of the fluid(s) should always be provided for the compressible version of turbulence models. One underlying reason is that the energy equation is always constructed and solved in the solvers where compressible turbulence models are used.

\subsection{Turbulence models for incompressible flows}

In OpenFOAM, the incompressible version for turbulence models are constructed by assuming that $\rho$ is constant. Thus, incompressible turbulence models in OpenFOAM are actually designed for strict incompressible flows. For instance, the corresponding incompressible version of Eq. \eqref{eq:kEquation_full} reads 
\begin{equation}
\label{eq:kEquation_const}
\frac{\partial k}{\partial t}+ \nabla \cdot (\vec{u} k)= P - \epsilon +\nabla \cdot \left[ \left(\nu + \frac{\nu_t}{\sigma_k}\right) \nabla k \right],
\end{equation}
where $\nu$ and $\nu_t$ are kinematic viscosities corresponding to $\mu$ and $\mu_t$, respectively.

\begin{figure}[h!]
	\centering
	\includegraphics[width=0.5\linewidth]{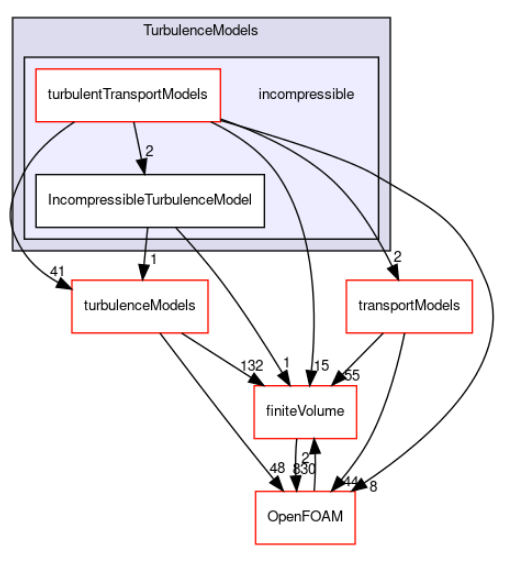}
	\caption{Directory dependency graph for incompressible turbulence models in OpenFOAM v1706 \citep{OpenCFD2018}. No thermal property is needed to construct incompressible turbulence models.}	
	\label{fig:incompressibleDependency}
\end{figure}

As shown in Fig. \ref{fig:incompressibleDependency}, the dependency graph for incompressible turbulence models is much simpler in comparison with the compressible one. One important difference is that the thermal properties of the fluid(s) are no longer needed to construct an incompressible turbulence model.

\section{Issues with turbulence modeling in isothermal VOF-based solvers}
\label{sect:issues}

The classification of incompressible turbulence models in OpenFOAM has a side effect on the turbulence modeling of variable-density incompressible flows. For such flows if the density change is caused by temperature, OpenFOAM has two solutions. One is using Boussinesq approximation for the momentum equation and utilizing the strict incompressible turbulence models. The other is giving up the divergence-free condition and using the full form of momentum equation and turbulence models directly. However, for isothermal VOF-based solvers, an inconsistency arises. Using the chain rule, Eq. \eqref{eq:kEquation_full} could be rewritten as
\begin{equation}
\label{eq:kEquation_r}
\left(\frac{\partial k}{\partial t}+ \nabla \cdot (\vec{u} k)\right)+
\frac {k}{\rho_m}\left(\frac{\partial \rho_m}{\partial t}+\vec{u} \cdot \nabla \rho_m\right)
= P - \epsilon + \nabla \cdot \left[ \left(\nu_m + \frac{\nu_t}{\sigma_k} \right) \nabla k \right] +\frac { \nabla \rho_m}{\rho_m} \cdot \left[ \left(\nu_m + \frac{\nu_t}{\sigma_k} \right) \nabla k \right].
\end{equation} 
In comparison with Eq. \eqref{eq:kEquation_const}, there is an additional term on the l.h.s. of Eq. \eqref{eq:kEquation_r}. According to the incompressible flow condition described by Eq. \eqref{eq:incompressible_rho}, this term should vanish. Therefore, Eq. \eqref{eq:kEquation_r} and Eq. \eqref{eq:kEquation_full} can both be rewritten into:
	\begin{equation}
	\label{eq:kEquation_rr}
	\frac{\partial k}{\partial t}+ \nabla \cdot (\vec{u} k)
	= P - \epsilon + \nabla \cdot \left[ \left(\nu_m + \frac{\nu_t}{\sigma_k} \right) \nabla k \right] +\frac { \nabla \rho_m}{\rho_m} \cdot \left[ \left(\nu_m + \frac{\nu_t}{\sigma_k} \right) \nabla k \right].
	\end{equation}	
In comparison with Eq. \eqref{eq:kEquation_const}, an extra term, which contains $\nabla \rho_m$, arises on the r.h.s. of Eq. \eqref{eq:kEquation_rr}. As long as $\rho_1 \neq \rho_2$ and $\nabla k \neq (0,0,0)$, this extra term is not zero for the transition zone. Therefore, the strict incompressible form of $k$ equation deviates from the original $k$ equation when it is applied to VOF simulations.

It is clear that this deviation is caused by the diffusion term where $\rho_m$ is inside the divergence operator. Even though the above derivation is based on Eq. \eqref{eq:kEquation_full}, the conclusion applies to governing equations for any variable $\phi$. As long as the diffusion term is in the form of $\nabla \cdot \left[ \left(\mu + \frac{\mu_t}{\sigma_{\phi}} \right) \nabla \phi \right]$, an inconsistency arises in the strict incompressible version when $\nabla \rho_m \neq (0,0,0)$ and $\nabla \phi \neq (0,0,0)$. A list of available turbulence models in the official release of OpenFOAM v1706, which are related to isothermal VOF simulations, is shown in Table \ref{table:models}. There are 6 models which are only available in strict incompressible form. Among all the other 24 models, which could be used in both compressible and strict incompressible forms, only 2 could avoid the deviation issue. 

\begin{table}[h!]
	\small
	\centering
	\caption{Turbulence models in OpenFOAM v1706}
	\label{table:models}
	\begin{tabular}{ccccc}
		\hline
		Turbulence models & Type
		&  \begin{tabular}{@{}c@{}} Available for   \\compressible flows\end{tabular}
		&  \begin{tabular}{@{}c@{}} Available for strict \\incompressible flows\end{tabular} 
		&  \begin{tabular}{@{}c@{}c@{}} Correct forms available\\ for variable-density \\incompressible flows\end{tabular} \\
		\hline
		SpalartAllmaras & RANS & $\checkmark$ &  $\checkmark$ & $\times$ \\
		kEpsilon &  RANS & $\checkmark$ &  $\checkmark$ & $\times$ \\
		RNGkEpsilon &  RANS & $\checkmark$ &  $\checkmark$ & $\times$ \\
		realizableKE &  RANS & $\checkmark$ &  $\checkmark$ & $\times$ \\
		LaunderSharmaKE & RANS &  $\checkmark$ &  $\checkmark$ & $\times$ \\
		kOmega &  RANS & $\checkmark$ &  $\checkmark$ & $\times$ \\
		kOmegaSST & RANS &  $\checkmark$ &  $\checkmark$ & $\times$ \\
		kOmegaSSTSAS & RANS &  $\checkmark$ &  $\checkmark$ & $\times$ \\
		kOmegaSSTLM  & RANS &  $\checkmark$ &  $\checkmark$ & $\times$ \\
		v2f & RANS &  $\checkmark$ &  $\checkmark$ & $\times$ \\
		LRR & RANS &  $\checkmark$ &  $\checkmark$ & $\times$ \\
		SSG & RANS &  $\checkmark$ &  $\checkmark$ & $\times$ \\
		qZeta & RANS &  $\times$ &  $\checkmark$ & - \\
		kkLOmega & RANS &  $\times$ &  $\checkmark$ & - \\
		LamBremhorstKE & RANS &  $\times$ &  $\checkmark$ & - \\
		LienLeschziner & RANS &  $\times$ &  $\checkmark$ & - \\
		ShihQuadraticKE & RANS &  $\times$ &  $\checkmark$ & - \\
		LienCubicKE & RANS &  $\times$ &  $\checkmark$ & - \\
		Smagorinsky & LES &  $\checkmark$ &  $\checkmark$ & $\checkmark$ \\
		WALE & LES &  $\checkmark$ &  $\checkmark$ & $\checkmark$ \\
		kEqn & LES &  $\checkmark$ &  $\checkmark$ & $\times$ \\
		dynamicKEqn & LES &  $\checkmark$ &  $\checkmark$ & $\times$ \\
		dynamicLagrangian & LES &  $\checkmark$ &  $\checkmark$ & $\times$ \\
		DeardorffDiffStress & LES &  $\checkmark$ &  $\checkmark$ & $\times$ \\
		SpalartAllmarasDES & DES &  $\checkmark$ &  $\checkmark$ & $\times$ \\
		SpalartAllmarasDDES & DDES &  $\checkmark$ &  $\checkmark$ & $\times$ \\
		SpalartAllmarasIDDES & IDDES &  $\checkmark$ &  $\checkmark$ & $\times$ \\
		kOmegaSSTDES & DES &  $\checkmark$ &  $\checkmark$ & $\times$ \\
		kOmegaSSTDDES & DDES &  $\checkmark$ &  $\checkmark$ & $\times$ \\
		kOmegaSSTIDDES & IDDES &  $\checkmark$ &  $\checkmark$ & $\times$ \\		
		\hline
		
	\end{tabular}
\end{table}

Mathematically, the strict incompressible turbulence models should not be applied to isothermal VOF simulations. However, these strict incompressible models are actually used in the corresponding solvers, e.g. \textit{interFoam}, \textit{interIsoFoam} and \textit{multiphaseInterFoam}. This issue, on the OpenFOAM side, is caused by the classification of turbulence models. However, the deeper-level reason is that, as mentioned in Section \ref{sect:introduction}, it is not easy to detect this issue due to the lack of reference values for turbulent multiphase flows. 

\section{Code development}
\label{sect:code}

Considering the popularity of both OpenFOAM and the VOF method, we are devoted to providing the community new open-source isothermal VOF-based solvers for turbulent multiphase flow investigations, of which the most basic and important feature is that these solvers will be able to use variable-density incompressible turbulence models. In this section, we illustrate our coding considerations for both users and developers, evaluate various possible solutions, implement the code and finally verify our implementation.

\subsection{Usability}

The new solvers are developed with the hope to replace corresponding existing solvers in OpenFOAM. Therefore, they are designed for common users who are unnecessarily able to write their solvers or even conscious of the issues stressed in Section \ref{sect:issues}. To the users, the usage of the new solvers should be as similar to the existing solvers as possible. Ideally, input files used for the new solvers should be exactly the same to those for the existing solvers.

\subsection{Object-oriented programming}

In order to make our solution developer-friendly, the object-oriented programming should be used. From this point of view, the new solvers should reuse as many existing codes as possible and make the minimum changes to where it is really needed. In order to reuse the code, a brief introduction on how the existing code is written is firstly given below. Two possible approaches of constructing variable-density incompressible turbulence models are then provided.

\subsubsection{Coding strategy for turbulence models in OpenFOAM}
\label{sect:strategy}

As mentioned in Section \ref{sect:vof}, the reason for assuming that the flow is incompressible is to utilize a simplified form of governing equations and to enhance the solver performance. For instance, all the 6 models, as listed in Table \ref{table:models}, which are only available in strict incompressible forms, are constructed based on the constant-density assumption. Take the LienCubicKE model for instance, as shown in Listing \ref{lst:LienCubicKE}, the $k$ equation is constructed exactly according to Eq. \eqref{eq:kEquation_const}, where $\rho$ is not involved. However, the disadvantage is also quite obvious that these simplified models are no longer valid for compressible flows. 
\begin{lstlisting}[caption = { $k$ equation for LienCubicKE model in Table \ref{table:models}.}, label = {lst:LienCubicKE}]
tmp<fvScalarMatrix> kEqn
(
    fvm::ddt(k_)
  + fvm::div(phi_, k_)
  - fvm::laplacian(DkEff(), k_)
  ==
    G //Production (Generation) term
  - fvm::Sp(epsilon_/k_, k_)
);
\end{lstlisting}
For each one of all the other 24 turbulence models listed in Table \ref{table:models}, there is only one template class, which could be used to construct corresponding turbulence models. Take the kEpsilon model for instance, as shown in Listing \ref{lst:kEpsilon}, the $k$ equation is constructed based on Eq. \eqref{eq:kEquation_full}. It should be noted that \texttt{alpha} in Listing \ref{lst:kEpsilon} is not the $\alpha$ in Eq. \eqref{eq:VOF}. When the multi-fluid approach, where each fluid has its own governing equations, is used for multiphase modeling, \texttt{alpha} means the volumetric fraction of a given fluid. This means that the kEpsilon model could even be used in a multi-fluid form, not only the compressible and strict incompressible forms shown in Table \ref{table:models}. All these forms are just different objects of the same template class. However, for flows with only one set of governing equations, e.g. single-phase flows and multiphase flows in the framework of VOF, \texttt{alpha} is simply unity.
  
\begin{lstlisting}[caption = { $k$ equation for kEpsilon model in Table \ref{table:models}.}, label = {lst:kEpsilon}]
tmp<fvScalarMatrix> kEqn
(
  fvm::ddt(alpha, rho, k_)
+ fvm::div(alphaRhoPhi, k_)
- fvm::laplacian(alpha*rho*DkEff(), k_)
==
  alpha()*rho()*G
- fvm::SuSp((2.0/3.0)*alpha()*rho()*divU, k_) //Compressibility contribution
- fvm::Sp(alpha()*rho()*epsilon_()/k_(), k_)
+ kSource()
+ fvOptions(alpha, rho, k_)
);
\end{lstlisting}
Listing \ref{lst:constructIncompressible} shows how strict incompressible forms are constructed from their corresponding template classes. Both \texttt{alpha} and \texttt{rho} in Listing \ref{lst:kEpsilon} are set to unity according to the above discussion. However, \texttt{divU} in Listing \ref{lst:kEpsilon}, which stands for $\nabla \cdot \vec{u}$, is still calculated even though it is zero according to the divergence-free condition. Solving equations similar to that in Listing \ref{lst:kEpsilon} is definitely slower than slowing equations like that in Listing \ref{lst:LienCubicKE}. However, this is the coding strategy of OpenFOAM due to the following two reasons. On the one hand, the most time-consuming part of the simulation is solving the Poisson's equation for pressure, and it is quite cheap to solve equations for turbulent quantities. On the other hand, this strategy significantly increases the reusability and maintainability of the codes since we only need to take care of those template classes, not separate turbulence models.

\begin{lstlisting}[caption = {Create strict incompressible turbulence models from template classes.}, label = {lst:constructIncompressible}]
makeBaseTurbulenceModel
(
geometricOneField, //alpha is unity
geometricOneField, //the density is also unity
incompressibleTurbulenceModel,
IncompressibleTurbulenceModel,
transportModel
);
\end{lstlisting}

\subsubsection{Brute-force approach}
The desired variable-density incompressible turbulence models could be constructed by a brute-force approach, which refers to any approach that works on individual turbulence models. This might be an easy and efficient way if we only need to modify few models. In fact, there are several alternatives for the brute-force approach, as described in \ref{apdx:alternatives}. However, it becomes infeasible when there are tens of models involved due to the following reasons. First, for a given model, we need to change all the equations, and the number is usually larger than one. Second, we have to give new names to all the newly created models, otherwise they will overwrite the original models. However, this is not compatible with the OpenFOAM naming convention and dramatically increases the number of models. As a result, the burden on code managements overweights the fact that only a few solvers benefit from this change. In addition, since it is unlikely that this approach will be adopted in the official release, the users have to compile all these models into other libraries and manually load these libraries when they want to use these models. Considering all these aspects, this brute-force approach is not adopted in the present study, of which the goal is to benefit common users.

\subsubsection{Object-oriented approach}
The present work uses an object-oriented approach to construct the desired turbulence models, which will be discussed in detail in Section \ref{sect:implementation}. This approach works on higher levels of the codes, which keeps the existing turbulence model template classes untouched. Therefore, the maintainability of codes is guaranteed and the naming convention of OpenFOAM is preserved as well. This benefits both the user and the developer.

\subsection{Code implementation}
\label{sect:implementation}

A step-by-step introduction is provided below to show how the new solvers are created.

\subsubsection{New class for incompressible turbulence models with varying density}

A new class, \texttt{varRhoIncompressibleTurbulenceModel}, is created for the isothermal turbulence models where the density, \texttt{rho}, is explicitly referenced in its constructor, as shown in Listing \ref{lst:constructor}. We use the prefix \texttt{varRho} to denote that this class is designed for variable-density flows. The same prefix will be used for the newly designed solvers as well. 

\begin{lstlisting}[caption = {Constructor for \textit{varRhoIncompressibleTurbulenceModel}.}, label = {lst:constructor}]
Foam::varRhoIncompressibleTurbulenceModel::varRhoIncompressibleTurbulenceModel
(
const volScalarField& rho,
const volVectorField& U,
const surfaceScalarField& alphaRhoPhi,
const surfaceScalarField& phi,
const word& propertiesName
)
:
turbulenceModel
(
U,
alphaRhoPhi,
phi,
propertiesName
),
rho_(rho)
{}
\end{lstlisting}

\subsubsection{Construct models from existing turbulence model template classes}

With the new class, all the 24 full-form turbulence models listed in Table \ref{table:models} could be constructed without reading thermal properties from the input files, as shown in Listing \ref{lst:constructNew}. For any customized turbulence model, as long as the full-form governing equations are available, this new class could create a corresponding object for variable-density flows as well.

\begin{lstlisting}[caption = {Create variable-density incompressible turbulence models from template classes..}, label = {lst:constructNew}]
makeBaseTurbulenceModel
(
geometricOneField,
volScalarField, //the varing density is to be read into this volScalarField
vIncompressibleTurbulenceModel,
VIncompressibleTurbulenceModel,
transportModel
);
\end{lstlisting}

\subsubsection{New solvers for isothermal VOF-based flows}

In varRhoTurbVOF, several isothermal VOF-related solvers are provided with which the full-form turbulence models are employed. All these solvers are modified based on the corresponding solvers in the official release of OpenFOAM. Since almost the same changes are made to each of these solvers, only one example is given here to illustrate how to change the existing \textit{interIsoFoam} solver to the newly designed \textit{varRhoInterIsoFoam}. It should be mentioned that such modifications also apply to isothermal VOF-based solvers in other versions of OpenFOAM and user-customized isothermal VOF-based solvers.

The first step is to modify the preprocessor directives in the main file of the solver. In \textit{interIsoFoam}, \texttt{turbulentTransportModel.H} is included for the construction of the strict incompressible turbulence models, as shown in Listing \ref{lst:interIsoFoam}. This should be substituted with \texttt{varRhoTurbulentTransportModel.H} in \textit{varRhoInterIsoFoam} such that the full form of turbulence models, where the density is explicitly included, could be used, as shown in Listing \ref{lst:vInterIsoFoam}.

\begin{lstlisting}[caption = {Preprocessor directives in \texttt{interIsoFoam.C}.}, label = {lst:interIsoFoam}]
#include "isoAdvection.H"
#include "fvCFD.H"
#include "subCycle.H"
#include "immiscibleIncompressibleTwoPhaseMixture.H"
#include "turbulentTransportModel.H"
#include "pimpleControl.H"
#include "fvOptions.H"
#include "CorrectPhi.H"
\end{lstlisting}

\begin{lstlisting}[caption = {Preprocessor directives in \texttt{varRhoInterIsoFoam.C}.}, label = {lst:vInterIsoFoam}]
#include "isoAdvection.H"
#include "fvCFD.H"
#include "subCycle.H"
#include "immiscibleIncompressibleTwoPhaseMixture.H"
#include "varRhoTurbulentTransportModel.H"
#include "pimpleControl.H"
#include "fvOptions.H"
#include "CorrectPhi.H"
\end{lstlisting}

The second step is to change the part which actually constructs the turbulence model. As shown in Listing \ref{lst:interIsoFoamCreateFields}, \textit{interIsoFoam} only needs the velocity field \texttt{U} and the flux field \texttt{phi} to construct the turbulence field, and the density \texttt{rho} is not included in \texttt{phi}. As for \textit{varRhoInterIsoFoam}, two additional fields, i.e. \texttt{rho} and \texttt{rhoPhi}, are necessary for turbulence model construction, as shown in Listing \ref{lst:vInterIsoFoamCreateFields}. It should be noted that both \texttt{rhoPhi} and \texttt{phi} are available in the isothermal VOF solvers, explicitly taking \texttt{phi} as an argument enhances the performance of the turbulence models.

\begin{lstlisting}[caption = {Turbulence model construction in \textit{interIsoFoam}.}, label = {lst:interIsoFoamCreateFields}]
autoPtr<incompressible::turbulenceModel> turbulence
(
incompressible::turbulenceModel::New(U, phi, mixture)
);
\end{lstlisting}

\begin{lstlisting}[caption = {Turbulence model construction in \textit{varRhoInterIsoFoam}.}, label = {lst:vInterIsoFoamCreateFields}]
autoPtr<incompressible::turbulenceModel> turbulence
(
incompressible::turbulenceModel::New(rho, U, rhoPhi, phi, mixture)
);
\end{lstlisting}

Only these two changes are needed to enable the new solver to use the full-form turbulence models. Therefore, all the other features of the existing solvers are preserved.

\subsection{Verification}

In order to confirm the correctness of the code implementation, verification tests are carried out with carefully selected test cases. Different solvers, e.g. two-phase and multiphase, various turbulence models, e.g. RANS and LES, and diverse features, e.g. adaptive mesh refinement and parallel computation, are all covered by the selected test cases. It should be noted that the validation of any specific turbulence model is out of the scope of the verification test and the scope of the present study.

The first stage is the consistency verification, which is based on the special case shown in Table \ref{table:models} that both Smagorinsky and WALE models could be correctly constructed in the official release. This implies that when these two models are used, both new and original solvers should provide the same result, which is verified in \ref{apdx:consistency}.

The second stage is the modification verification, which aims at verifying that changes have been made to the codes, as detailed in \ref{apdx:modification}. As mentioned in Section \ref{sect:introduction}, due to the lack of reference values, it is still not verified whether such changes are correctly implemented according to the corresponding governing equations.

The third stage is the cross verification, which is a good practice in the absence of reference values. We have shown several alternative implementations for any given turbulence model in \ref{apdx:alternatives}. A cross verification between the object-oriented approach and other alternatives is provided in \ref{apdx:cross}. This cross verification not only proves the correctness of the code implementation, but also shows the numerical stability of the implemented code.

\section{Usage}
\label{sect:usage}
The source code is provided in \citep{Fan2018d2}. In order to use the code, one needs to load the environment variable for OpenFOAM v1706 first and then run \texttt{./compile.sh} to compile the code. As for the usage for a specific solver, e.g. \textit{varRhoInterFoam}, it is almost the same with the corresponding existing solver \textit{interFoam}. For instance, it could be executed on 1024 processors by simply typing \texttt{mpirun -np 1024 varRhoInterFoam -parallel} in the terminal. 

Since the full-form turbulence models are used in the new solvers, the corresponding discretization schemes should be provided to solve the governing equations numerically. Other than this, the users could reuse all their $interFoam$ input files for $varRhoInterFoam$.

\section{Performance evaluation}
\label{sect:performance}
With both the new and original solvers available, we conduct a simple performance evaluation based on the experiment conducted by \citep{Fabre1987}. The comparison will justify our motivation for developing the new solvers.

The experiments were carried out in a rectangular flow channel with a 0.1\% downward slope. The channel was 12.6 m long, 20 cm wide and 10 cm high. Three co-current air-water stratified flows are investigated in the present study and the corresponding flow configurations are listed in Table \ref{table:flow}.

\begin{table}[h!]
	\centering
	\caption{Flow configurations in the experiment \citep{Fabre1987}.}
	\label{table:flow}
	\begin{tabular}{ccc}
		\hline
		Run reference & Water flow rate [L/s] & Air flow rate [L/s] \\
		\hline
		250 & 3.0 & 45.4 \\
		400 & 3.0 & 75.4 \\
		600 & 3.0 & 118.7 \\
		\hline
	\end{tabular}
\end{table}

\begin{figure}[h!]
	\centering
	\includegraphics[width=0.9\linewidth]{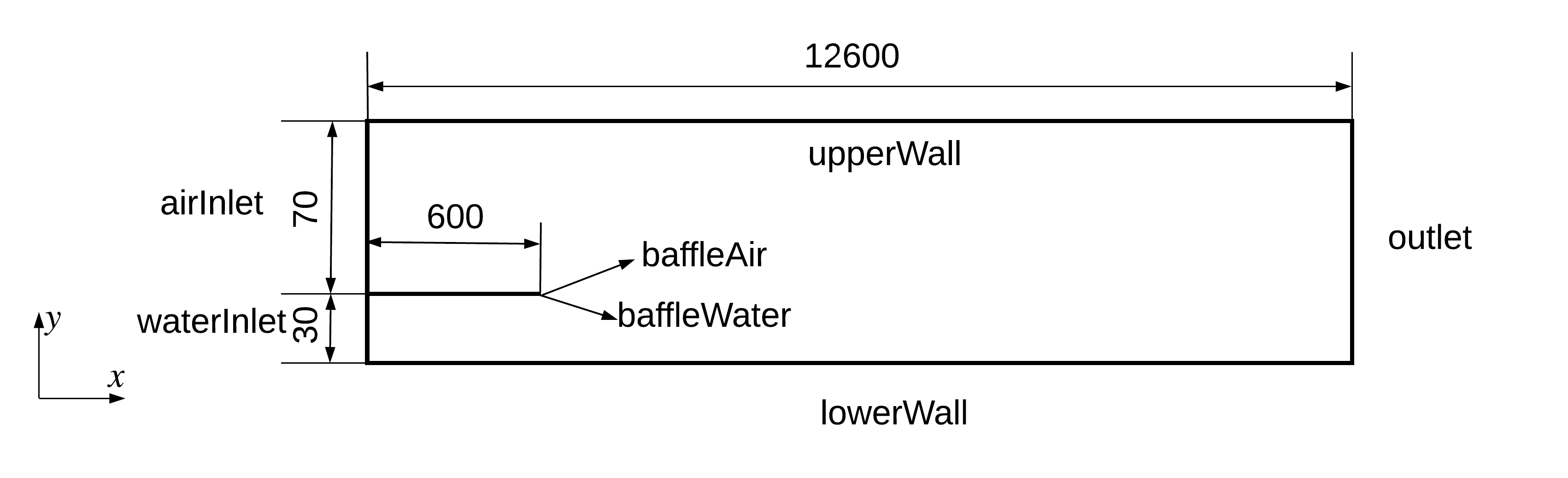}
	\caption{Sketch of the computational domain (unit in mm, not to scale).}	
	\label{fig:computationalDomain}
\end{figure}

A 2D computational domain is constructed as shown in Fig. \ref{fig:computationalDomain}. Similarly to the experiment, air and water are supplied via corresponding inlets. These two inlets are assumed to be separated by a zero-thickness 100 mm-long baffle. One reason for making such assumption is that details of the baffle are not provided in the paper. Another reason is that the measuring zone is quite far away from the inlets indicating that the detailed inlet configurations of the inlet region should only have minor effects on the results of the measuring zone.

\subsection{Boundary conditions}

Boundary conditions for all the flow variables are listed in Table \ref{table:boundary conditions}.
\begin{small}
\begin{center}	
	\begin{threeparttable}[h!]
		
		\caption{Boundary conditions.}
		\label{table:boundary conditions}
		\begin{tabular}{cccccccc}
			\hline
			& airInlet & waterInlet & outlet & upperWall & lowerWall & baffleAir & baffleWater\\
			\hline
			$\alpha$ & $\alpha = 0$ & $\alpha = 1$ & $\nabla \alpha = 0$ & $\alpha = 0$ & $\alpha = 1$ & $\alpha = 0$ & $\alpha = 1$\\
			$\vec{u}$ & mappedC \tnote{*} & mappedC & advective & no slip & no slip  & no slip & no slip\\
			$p_{rgh}$ & fixed flux & fixed flux &  \begin{tabular}{@{}c@{}} fixed total \\pressure\end{tabular} & fixed flux & fixed flux & fixed flux & fixed flux\\
			$k$ & mappedN \tnote{**} & mappedN & $\nabla k = 0$ & $k = 0$ & $k = 0$ & wall function & wall function\\
			$\omega$ & mappedN & mappedN & $\nabla \omega = 0$ & wall function\tnote{***} & wall function & wall function & wall function\\			
			\hline
		\end{tabular}		
		\begin{tablenotes}
			\item[*] mapped condition with the constraint on the average value.
			\item[**] mapped condition without constraints.
			\item[***]there is a bug in $omegaWallFunction$ in the official release of OpenFOAM v1706 and it is fixed in the present study.
		\end{tablenotes}		
		
	\end{threeparttable}	
\end{center}
\end{small}
\subsection{Results}

Three meshes with $\Delta y =$ 2 mm, 1 mm, and 0.5 mm are constructed for the simulations, where $\Delta y$ denotes the mesh size (around the interface) in the vertical direction. For each mesh, both the strict incompressible and variable-density incompressible version of kOmegaSST model are used for turbulence modeling. All the simulations are run in transient modes. After the initial-condition effects die out, the sampling is carried out for 50 s to consider the variations caused by the wavy interface.

\begin{figure}[h!]
	\centering
	\subfloat [pressure gradient \label{fig:dp250}]
	{\includegraphics [width=0.47\linewidth] {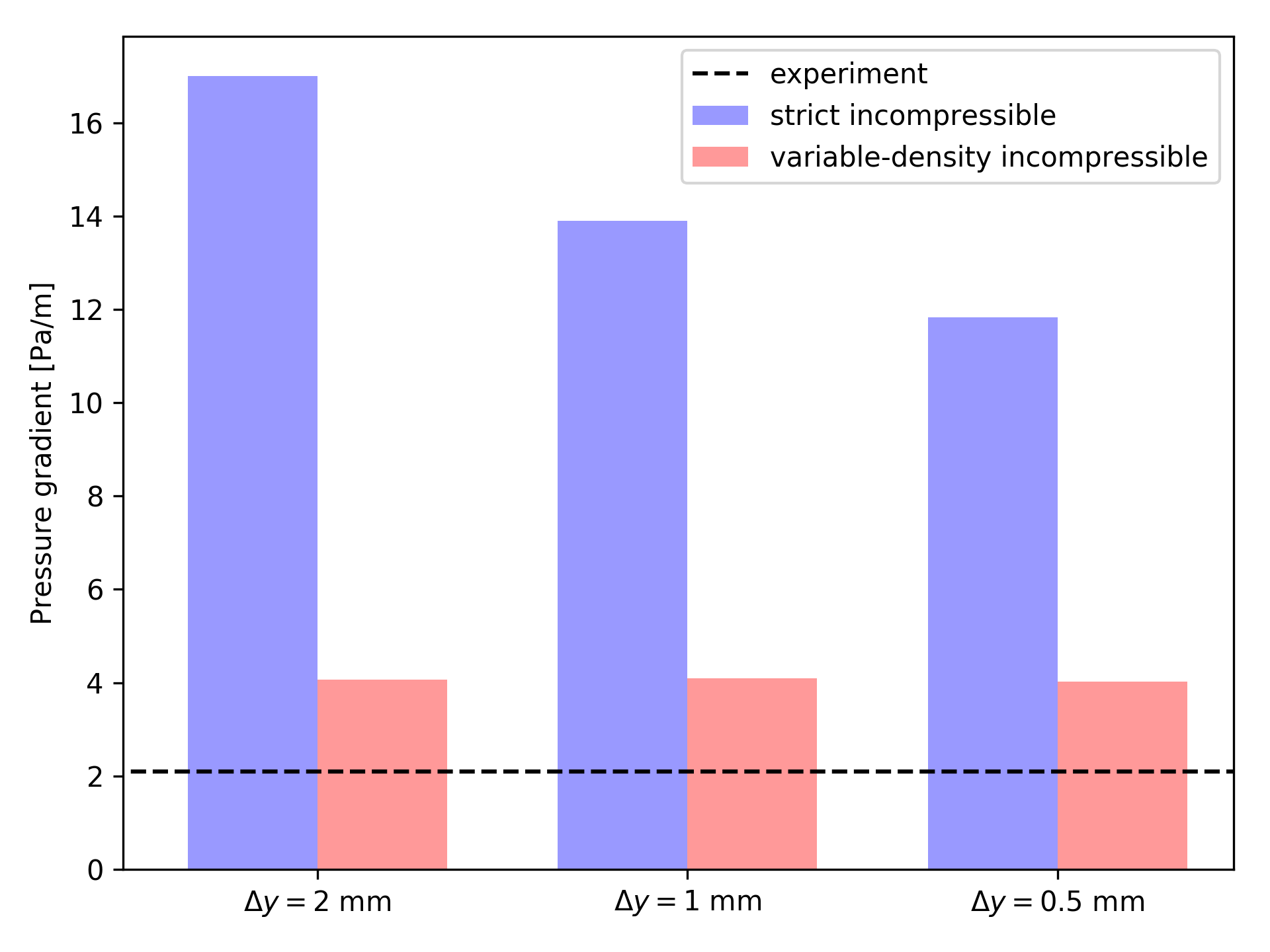}} 
	\subfloat[turbulent kinetic energy profiles in the vertical direction, SI: strict incompreissible, VI: variable-density incompressible \label{fig:k250}]{\includegraphics [width=0.525\linewidth] {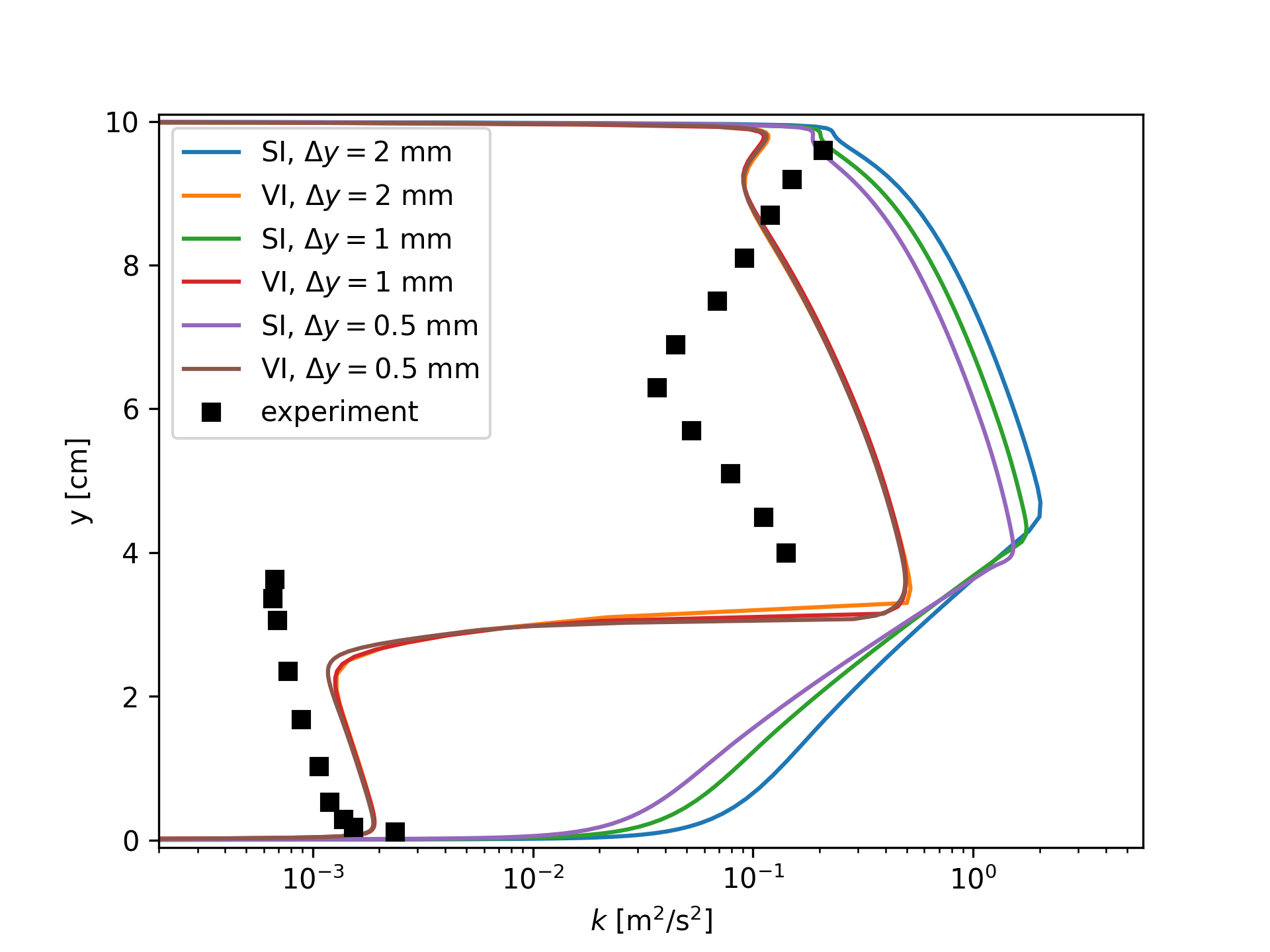}}
	\caption{Comparisons for pressure gradient and turbulent kinetic energy velocity profiles (run-250).}
	\label{fig:250}
\end{figure}

\begin{figure}[h!]
	\centering
	\subfloat [pressure gradient \label{fig:dp400}]
	{\includegraphics [width=0.47\linewidth] {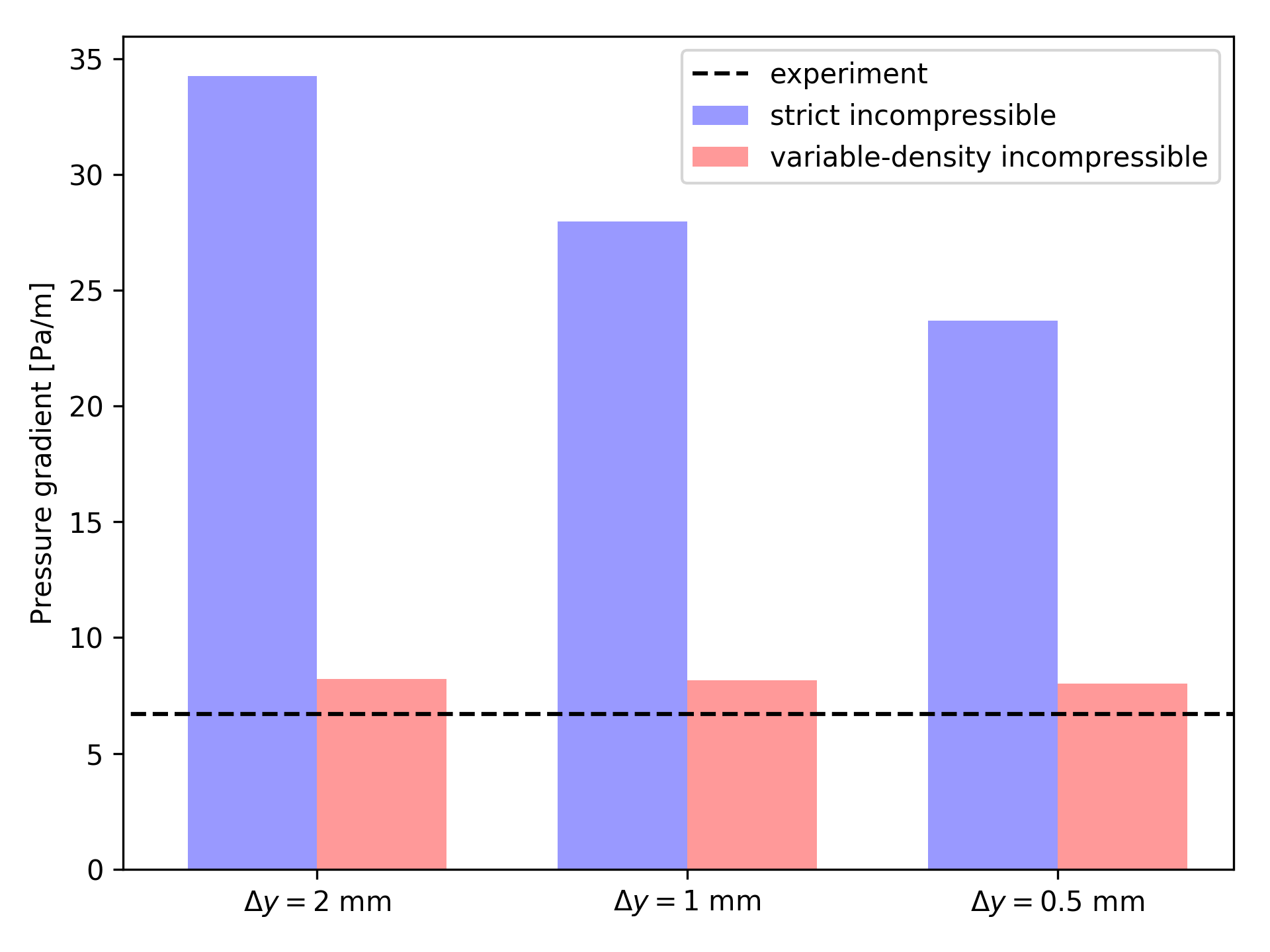}} 
	\subfloat[turbulent kinetic energy profiles in the vertical direction \label{fig:k400}]{\includegraphics [width=0.525\linewidth] {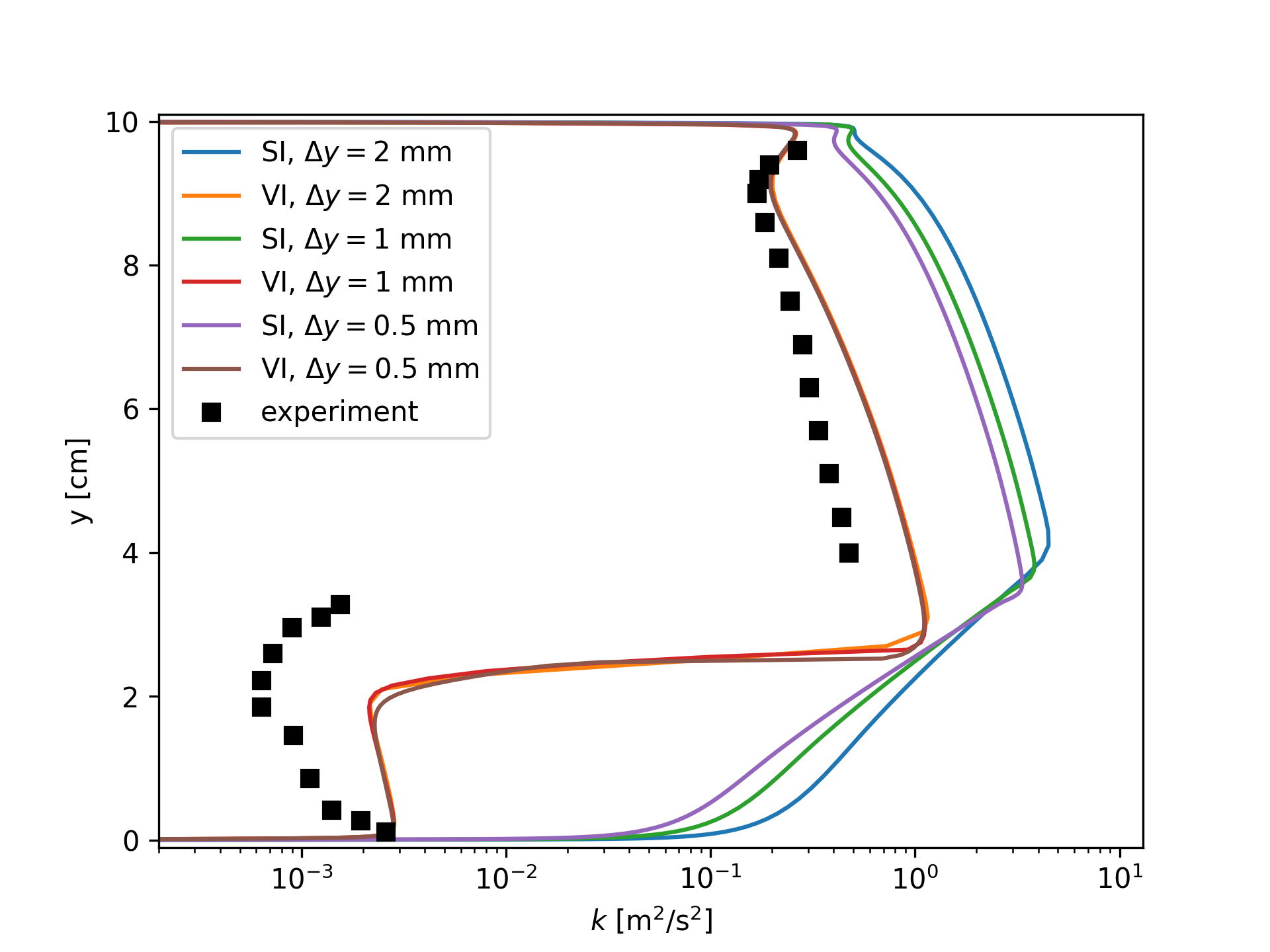}}
	\caption{Comparisons for pressure gradient and turbulent kinetic energy velocity profiles (run-400).}
	\label{fig:400}
\end{figure}

\begin{figure}[h!]
	\centering
	\subfloat [pressure gradient \label{fig:dp600}]
	{\includegraphics [width=0.47\linewidth] {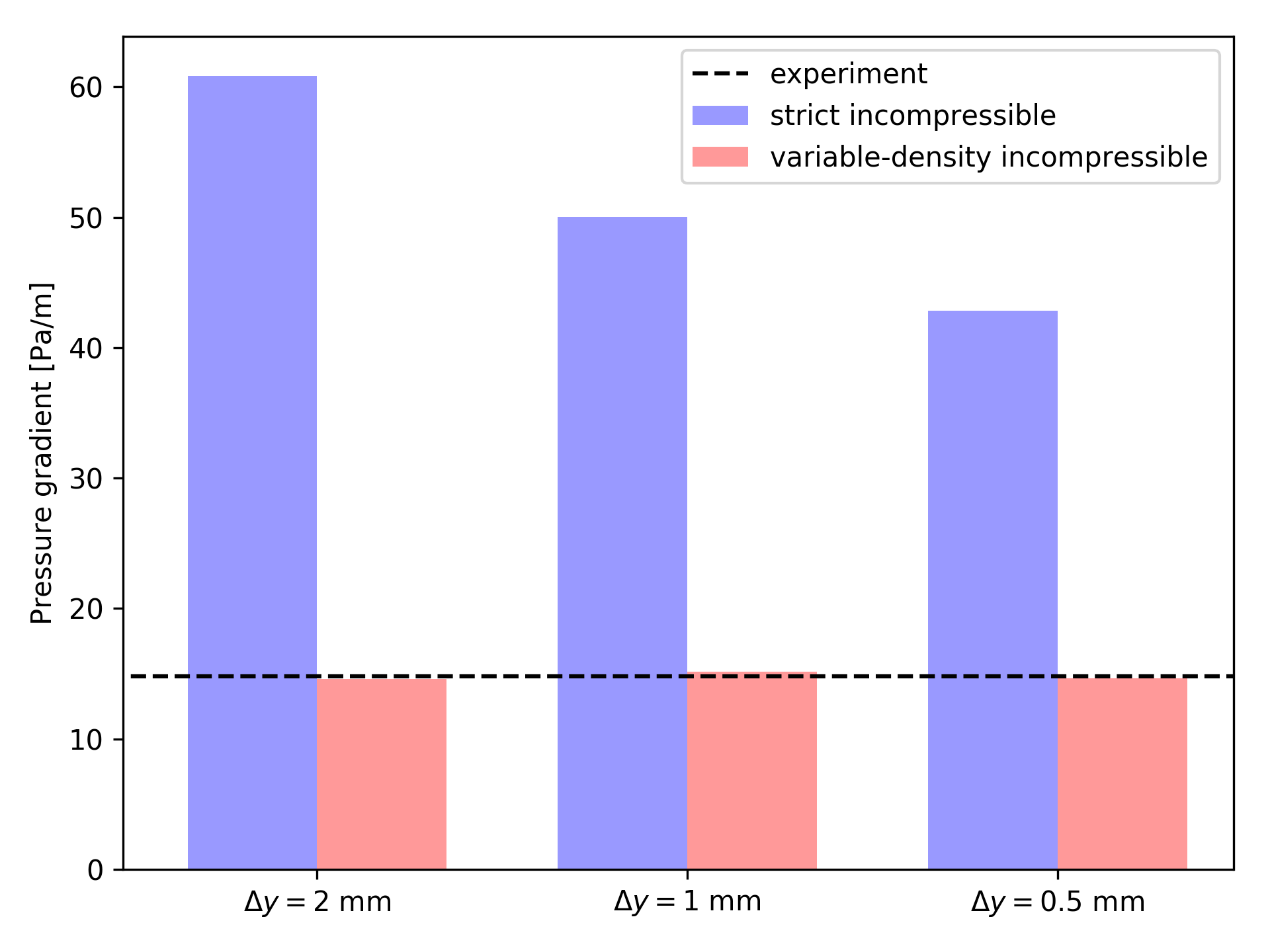}} 
	\subfloat[turbulent kinetic energy profiles in the vertical direction \label{fig:k600}]{\includegraphics [width=0.525\linewidth] {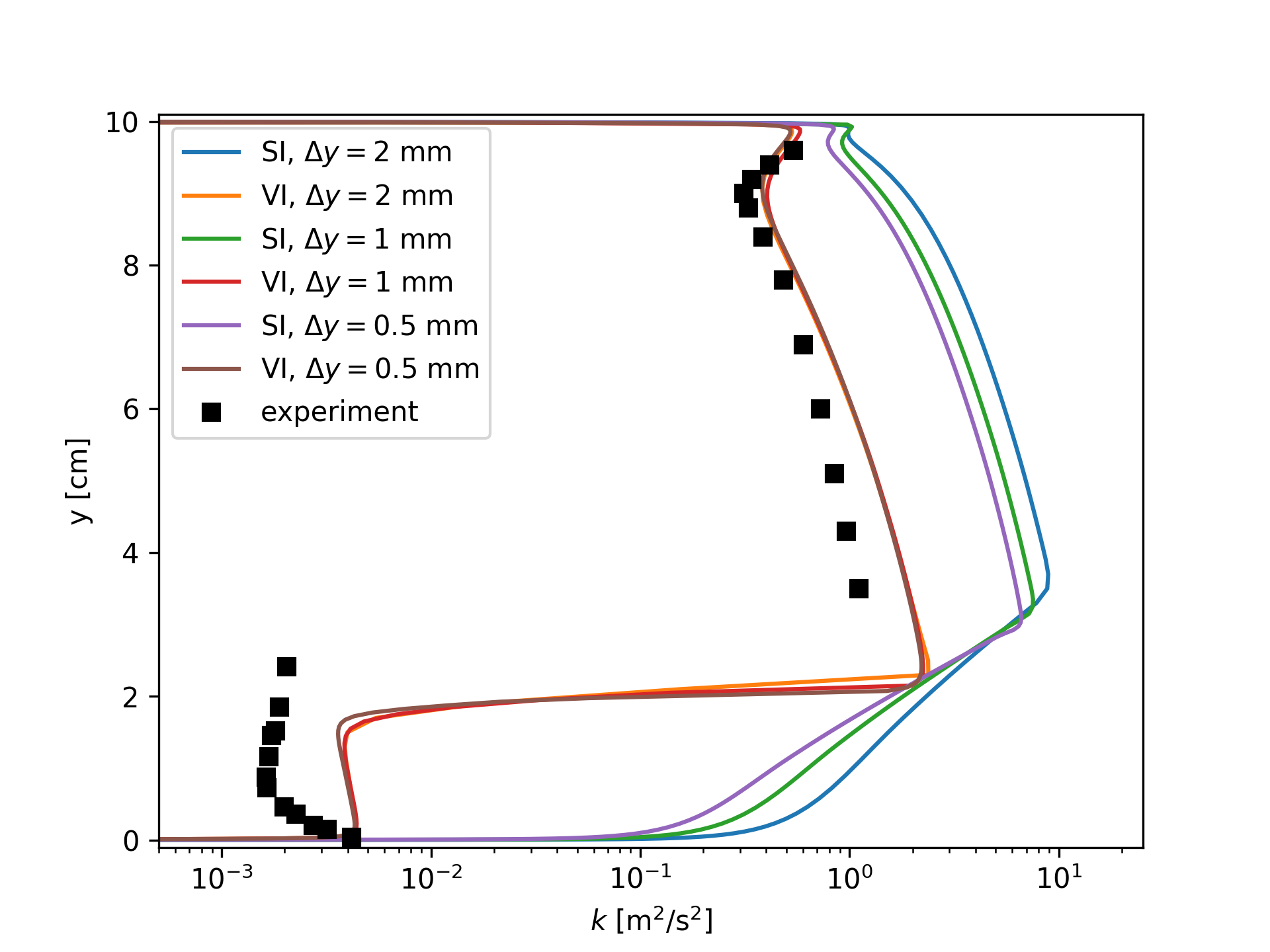}}
	\caption{Comparisons for pressure gradient and turbulent kinetic energy velocity profiles (run-600).}
	\label{fig:600}
\end{figure}

In Figs. \ref{fig:250}-\ref{fig:600}, the pressure drop and $k$ profiles in the fully developed regions are compared with the experimental values. Two important conclusions could be easily made from these figures. One is that the strict incompressible version is much more sensitive to mesh refinement. This makes it impractical to conduct sensitivity studies on the mesh size. On the other hand, the variable-density incompressible version could capture the abrupt change in $k$ around the interface. However, the strict incompressible version totally misses this abrupt change.

\begin{figure}[h!]
	\centering
	\subfloat [strict incompressilbe \label{fig:strict}]
	{\includegraphics [width=0.49\linewidth] {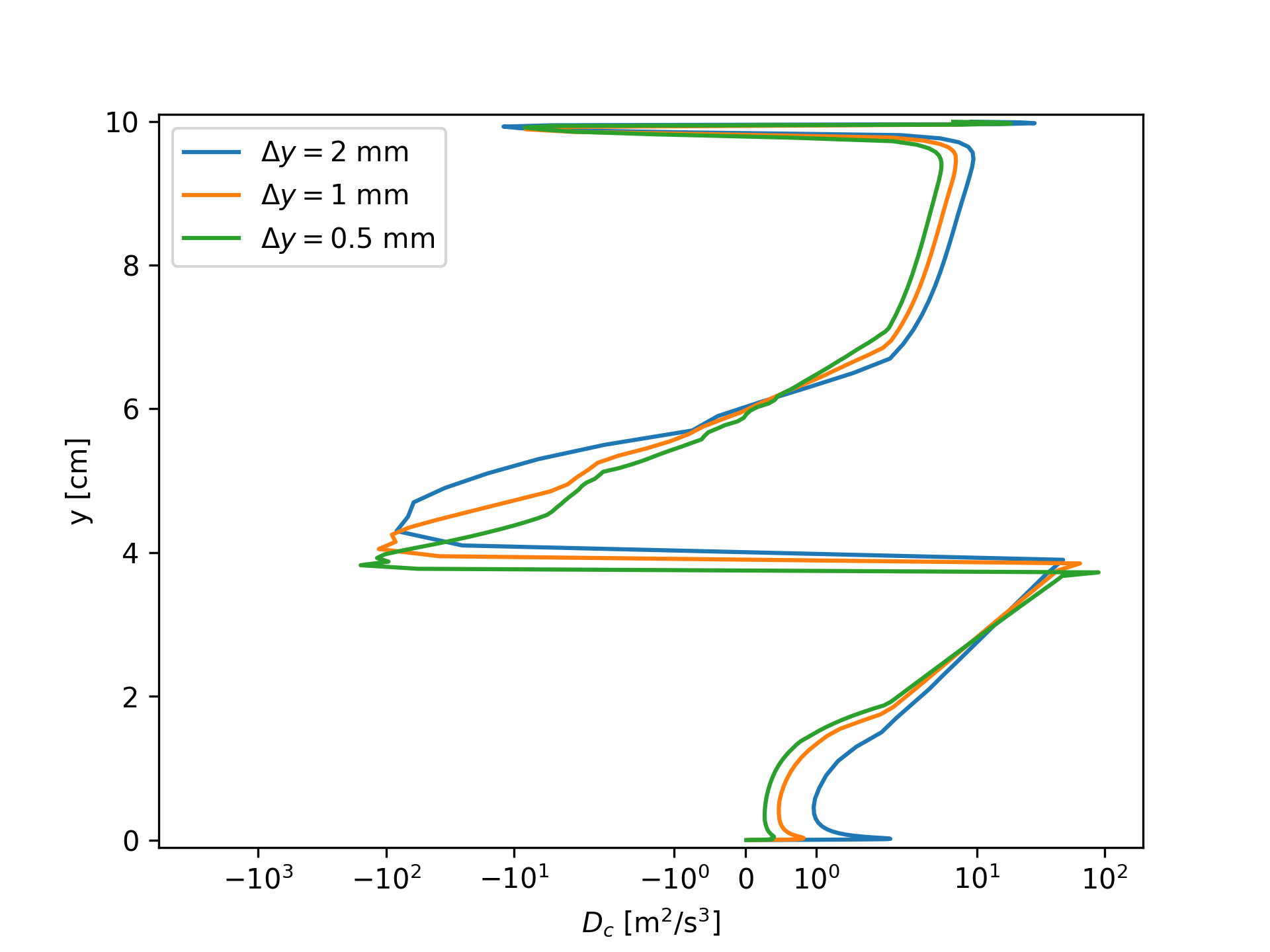}} 
	\subfloat[variable-density incompressible \label{fig:variable}]{\includegraphics [width=0.49\linewidth] {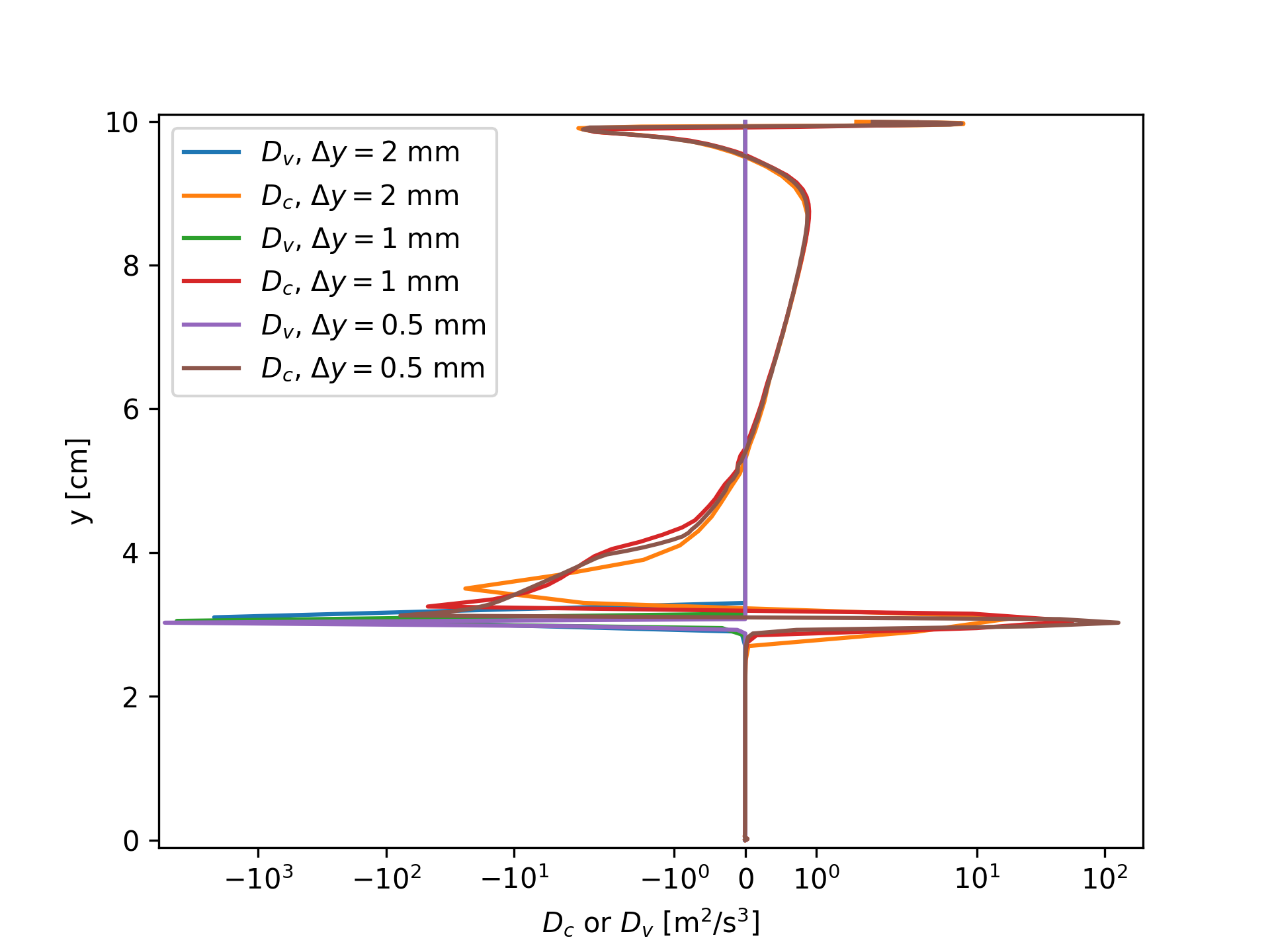}}
	\caption{Comparisons for diffusion terms for $k$ (run-250).}
	\label{fig:diffusion}
\end{figure}

According to the discussion in Section \ref{sect:issues}, this huge performance difference is caused by the fact that different diffusion terms are used in different models. We could define $D_c = \nabla \cdot \left[ \left(\nu_m + \frac{\nu_t}{\sigma_k} \right) \nabla k \right] $ as the constant-density part of the diffusion term and $D_v = \frac { \nabla \rho_m}{\rho_m} \cdot \left[ \left(\nu_m + \frac{\nu_t}{\sigma_k} \right) \nabla k \right]$ the variable-density part. For the strict incompressible turbulence modes, we only have the $D_c$ contribution, as shown in Fig. \ref{fig:strict}, and $D_c$ is positive around the interface. While for the variable-density models, two components both contribute, as shown in Fig. \ref{fig:variable}. There are two very important conclusions that we could get from Fig. \ref{fig:variable}. On the one hand, $D_v$ is negative around the interface. Therefore, strict incompressible turbulence models over-predict $k$ values around the interface due to the fact that $D_v$ is not included in the equation. On the other hand, in terms of the absolute value, $D_v$ is orders of magnitude higher than $D_c$ around the interface meaning that the total diffusion is dominated by $D_v$ for this region resulting in a negative total diffusion term around the interface. Therefore, strict incompressible turbulence models over-predict $k$ values to a significant extent around the interface, as already shown in Figs. \ref{fig:250}-\ref{fig:600}. More results are provided for run-250 in \ref{apdx:result}.

We are aware that the variable-density incompressible version still fails to match the experimental data on a quantitative level, and that turbulence damping proposed by \citep{Egorov2004} might help to give a better prediction. However, they are both out of the scope of the current study. Interested readers are referred to \cite{Fan2019} for more details. By conducting this performance evaluation, we are intending to show that the variable-density version could give a better prediction in comparison with the strict incompressible version. Therefore, the variable-density incompressible version of turbulence models should be used in VOF-based solvers.

\section{Conclusions and outlooks}
\label{sect:conclusion}
	Due to the limitations of the turbulence model classification in OpenFOAM, isothermal VOF-based solvers in the official release could only use the strict incompressible form of turbulence models, which is inconsistent with the fact that such solvers are intended to solve variable-density flows. 

	With the object-oriented paradigm, by making the minimal changes to the existing codes, the developed solvers could construct the correct turbulence models and, at the same time, preserve all the other features of existing solvers. All the newly designed solvers benefit from the new class for turbulence models. In addition, the newly designed class for isothermal variable-density turbulence models could also be applied to other flows that are not described by the VOF framework.
	
	The newly implemented solvers are released as open-source together with this paper \citep{Fan2018d2} with the hope that the solvers could be used and further tested against various flow conditions. Implementations for other recent OpenFOAM versions are provided in \cite{GitHub} (\url{https://github.com/wenyuan-fan/varRhoTurbVOF}). By providing the community a free and user-friendly access to various VOF solvers and turbulence models, we hope that turbulent VOF simulations could be widely conducted and further improved, and gradually catch up the demand from practical applications.
	
	\section*{Acknowledgement}
	
	The simulations were performed on resources provided by the Swedish National Infrastructure for Computing (SNIC) at PDC. Wenyuan Fan is also grateful for the support of China Scholarship Council (CSC).

	\appendix
	
\section{Candidates for the brute-force approach}
\label{apdx:alternatives}
The brute-force approach could be further divided into two groups, namely the source term approach and the full-form approach, depending on how existing codes are utilized in the newly created model.

\subsection{Source term approach}
The idea of this approach is quite simple and straightforward. According to the derivation in Section \ref{sect:issues}, Eq. \eqref{eq:kEquation_rr} only has one more term than Eq. \eqref{eq:kEquation_const}, and we could derive the exact form of this term for any given governing equation. Therefore, the obvious solution is adding this additional term, as shown in Eq. \eqref{eq:sourceTerm}, to the strict incompressible form to get the variable-density incompressible form. In this approach, all the other parts of the model could be inherited from the corresponding parent class. Therefore, code repetitions could be avoided. In terms of numerical realization for the gradient term, $\nabla k$, there are two methods available for the official release.
\begin{equation}
	\label{eq:sourceTerm}
	S_k = \frac { \nabla \rho_m}{\rho_m} \cdot \left[ \left(\nu_m + \frac{\nu_t}{\sigma_k} \right) \nabla k \right].
\end{equation}

\subsubsection{Source term with explicit gradient term}
$\nabla k$ could be calculated explicitly by using the existing value of $k$, which is denoted by the superscript \textit{old}, as shown in Eq. \eqref{eq:sourceTermEx}. Details of  implementing this term into the kEpsilon model could be found in \texttt{bruteForceExamples/VOFKEpsilonEx} folder of the source code. This might be the simplest brute-force approach to create a variable-density incompressible turbulence model. However, it has numerical stability issues due to the explicit treatment of the gradient term, as will be discussed in \ref{apdx:cross}.
\begin{equation}
\label{eq:sourceTermEx}
S_k = \frac { \nabla \rho_m}{\rho_m} \cdot \left[ \left(\nu_m + \frac{\nu_t}{\sigma_k} \right) \nabla k^{old} \right].
\end{equation}

\subsubsection{Source term with implicit gradient term}
$\nabla k$ could also be calculated implicitly by using the unknown value of $k$, which is denoted by the superscript \textit{new}, as shown in Eq. \eqref{eq:sourceTermEx}.
\begin{equation}
\label{eq:sourceTermIm}
S_k = \frac { \nabla \rho_m}{\rho_m} \cdot \left[ \left(\nu_m + \frac{\nu_t}{\sigma_k} \right) \nabla k^{new} \right].
\end{equation}
It should be noted that, in the official release of OpenFOAM, it is impossible to calculate Eq. \eqref{eq:sourceTermIm} directly due to the fact that the implicit gradient term is not available in matrix form. Since the implicit Laplacian term is available in matrix form, Eq. \eqref{eq:sourceTermIm} could be calculated using the following equivalent form which is derived from the chain rule
\begin{equation}
\label{eq:sourceTermImNew}
S_k = \frac { 1}{\rho_m} \nabla \cdot \left[ \rho_m \left(\nu_m + \frac{\nu_t}{\sigma_k} \right) \nabla k^{new} \right] - \nabla \cdot \left[ \left(\nu_m + \frac{\nu_t}{\sigma_k} \right) \nabla k^{new} \right].
\end{equation}
The details of adding this term into the kEpsilon model are provided in \texttt{{bruteForceExamples/VOFKEpsilonIm}} folder of the source code.
 
\subsection{Full-form approach}
This approach allows us to nominally construct a strict incompressible turbulence model but actually use a full-form model. The trick is that we still set the density field to unity when constructing the model. However, this density is a dummy field which will not be used in the turbulence model. Instead, the code is designed to be able to find out the real density and use it in the model. The similar procedure is applied to the flux field as well. The implementation details for the kEpsilon model are provided in \texttt{{bruteForceExamples/kEpsilonFull}} folder of the source code. In comparison with the source term approach, this method is more complicated since it needs to change the structure of existing turbulence models and introduces considerable code repetitions.

\section{Consistency verification}
\label{apdx:consistency}

\begin{figure}[h!]
	\centering
	\includegraphics[width=0.6\linewidth, clip = true, trim = 400 100 250 100]
	{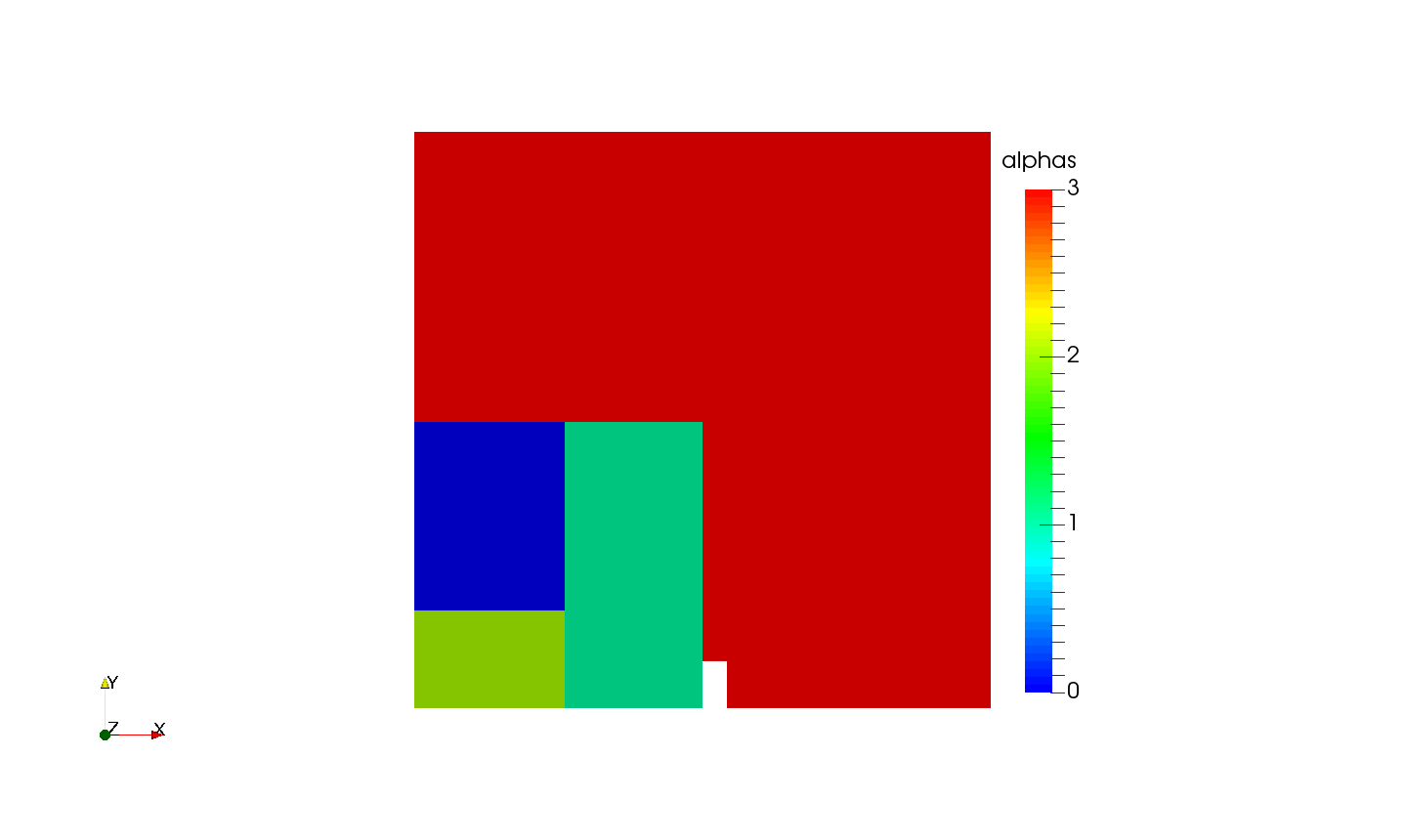}
	\caption{Initial condition for test case \textit{damBreak4Phase} where 0, 1, 2, 3 are used to denote water, oil, mercury and air, respectively.}	
	\label{fig:damBreak4Phase}
\end{figure}

\begin{figure}[h!]
	\centering
	\subfloat[\textit{multiphaseInterFoam}]{\includegraphics[width=0.49\linewidth, clip = true, trim = 400 100 400 100]
		{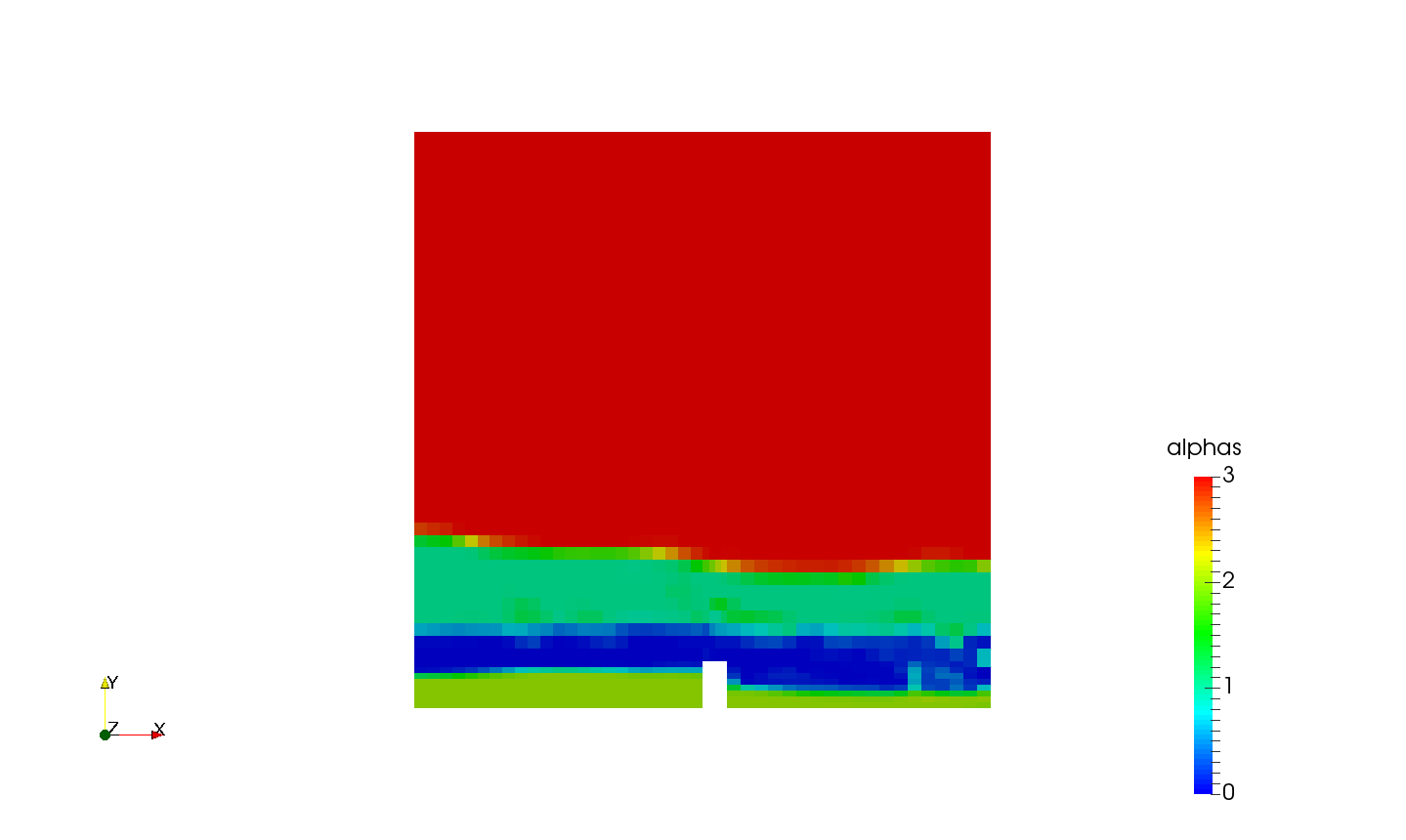}}
	\subfloat[\textit{varRhoMultiphaseInterFoam}]{\includegraphics[width=0.49\linewidth, clip = true, trim = 400 100 400 100]
		{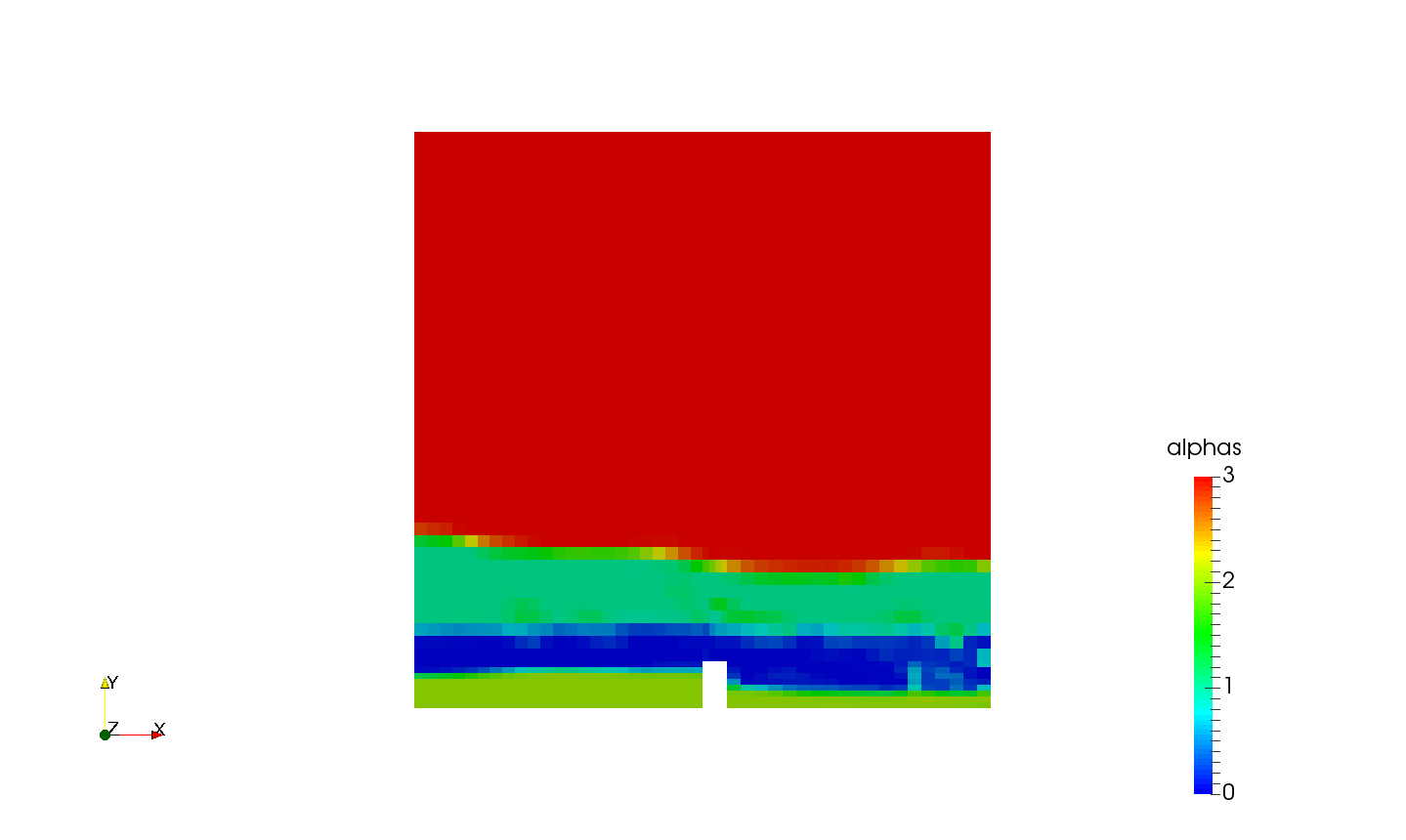}}
	
	\caption{Phase distribution at the end of simulations. Exactly the same result is predicted by both solvers with reasonable stratification caused by the density difference.}	
	\label{fig:damBreak4PhaseComparison}
\end{figure}

It is claimed in Table \ref{table:models} that both Smagorinsky and WALE models survive from the deviation issue which is caused by the classification of turbulence models in OpenFOAM. Therefore, both new and original solvers should provide exactly the same result when these two models are used. In order to verify this, the \textit{damBreak4Phase} tutorial for \textit{multiphaseInterFoam} is used for the consistency verification where the turbulence modeling is changed from laminar to LES with the Smagorinsky model. The initial condition is given in Fig. \ref{fig:damBreak4Phase}, where water, oil, mercury and air will change their positions as time proceeds due to the density differences. \textit{multiphaseInterFoam} and \textit{varRhoMultiphaseInterFoam} are used to run this case and the simulations last for 6 s. At the end of the simulation, both solvers give the same prediction for phase distributions, as shown in Fig. \ref{fig:damBreak4PhaseComparison}. This consistent performance of both new and original solvers justifies the statement that we have made in Table \ref{table:models} and also partially verifies the code implementation.

\section{Modification verification}
\label{apdx:modification}

\begin{figure}[h!]
	\centering
	\subfloat[\textit{interDyMFoam} \label{fig:motorbikea}]
	{\includegraphics[width=0.6\linewidth, clip = true, trim = 350 200 220 200]
		{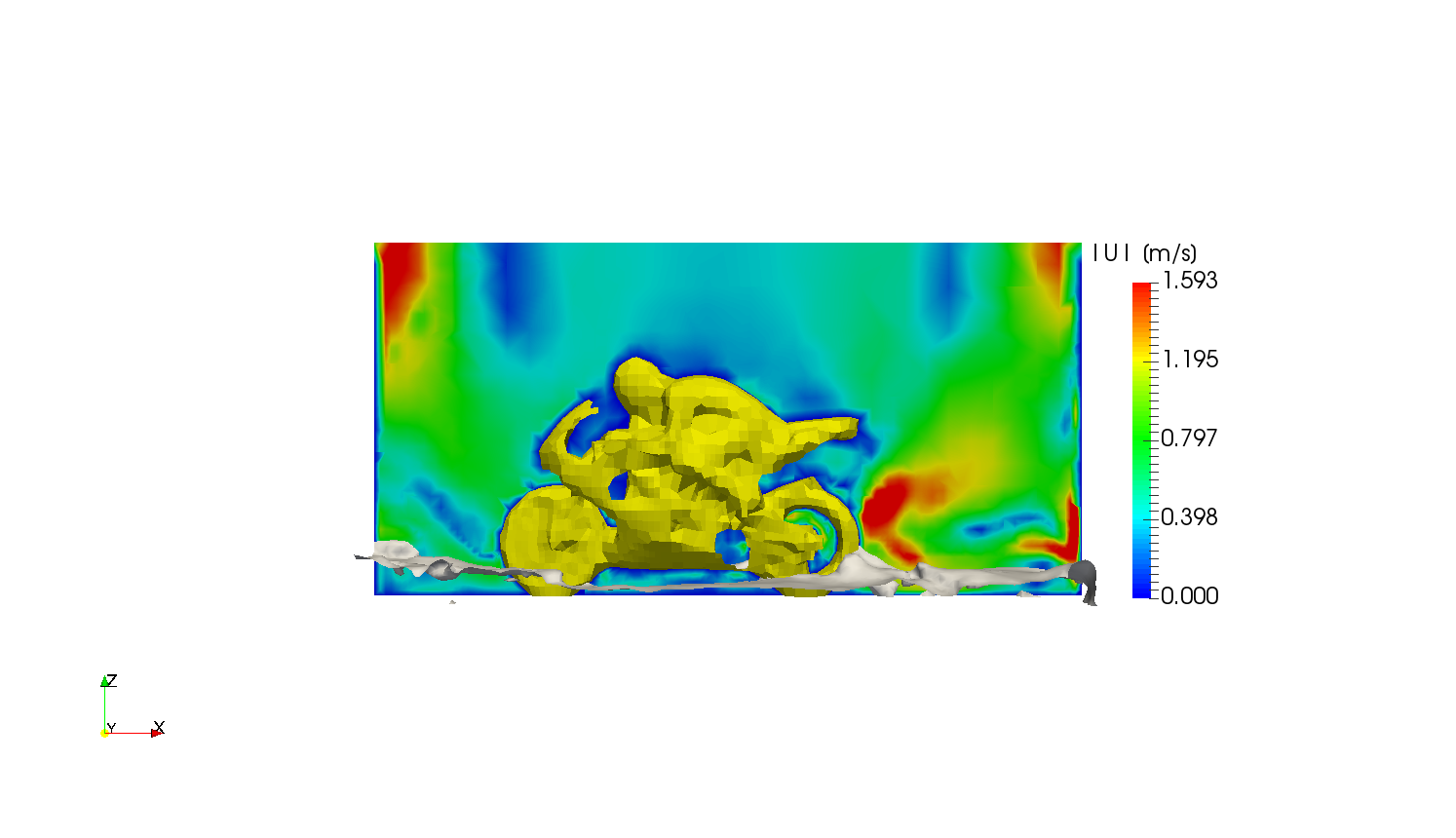}}
	
	\subfloat[\textit{varRhoInterDyMFoam}\label{fig:motorbikeb}] {\includegraphics[width=0.6\linewidth, clip = true, trim = 350 200 220 200]
		{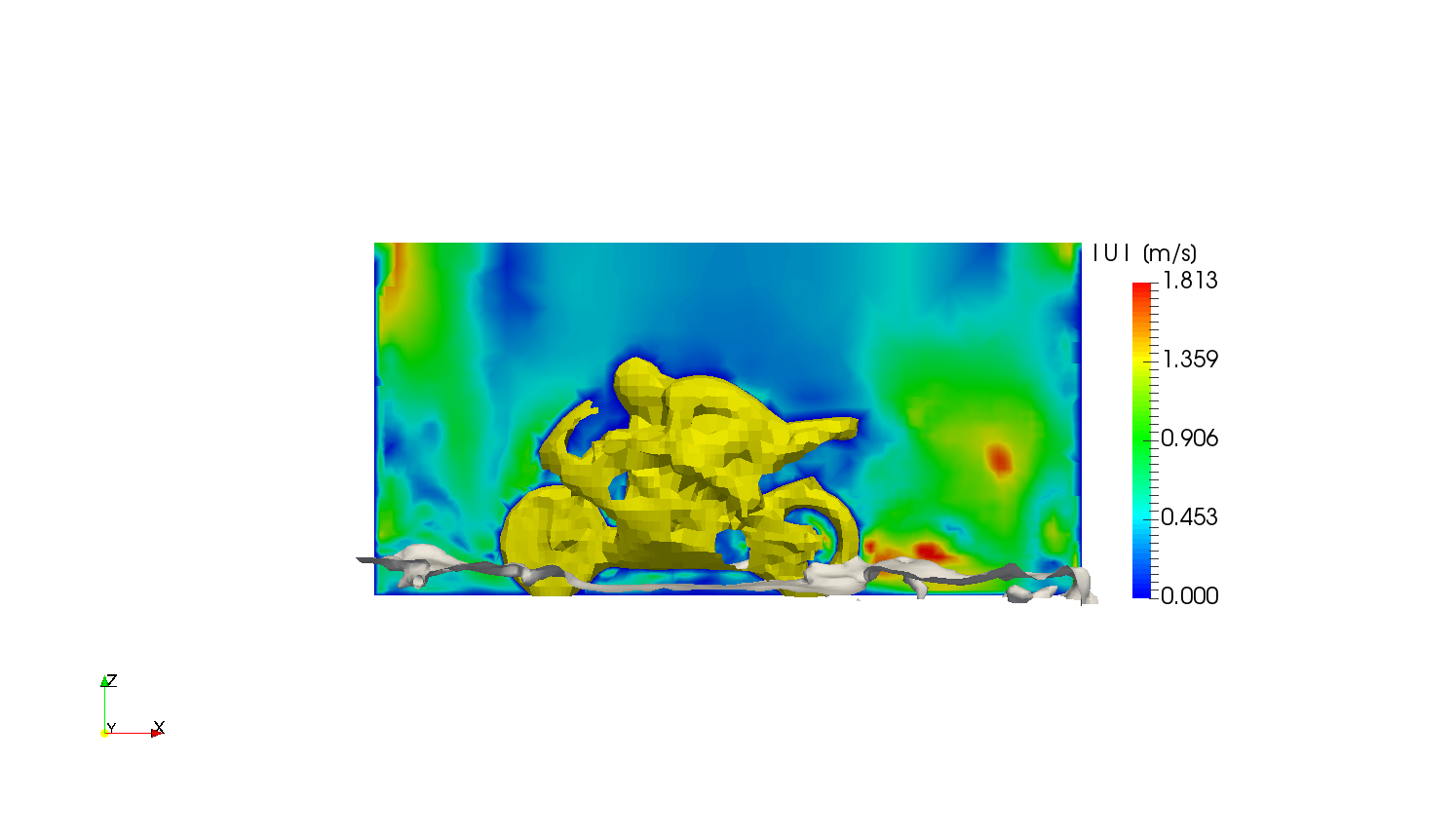}}
	
	\caption{Result comparison for \textit{interDyMFoam} and \textit{varRhoInterDyMFoam}: water-air interface is denoted by the gray surface; the middle plane of the domain is colored by the velocity magnitude.}	
	\label{fig:motorbike}
\end{figure}

The goal of the modification verification is to show that a new class of turbulence models are constructed in the new solvers. The \textit{motorBike} case for \textit{interDyMFoam} is used for this test. In the simulation, a motorbike runs on the ground which is covered by water. The flow is turbulent and a turbulence model (kOmegaSST in Table \ref{table:models}) is used in the simulations. As shown in Fig. \ref{fig:motorbike}, two solvers give quite different predictions for the shape of water-air interface and the distribution of velocity magnitude. This proves that \textit{varRhoInterDyMFoam} does use a different turbulence model when compared with \textit{interDyMFoam}. However, further verifications are still needed to confirm that such changes are correctly implemented. 

\newpage
\section{Cross verification}
\label{apdx:cross}
\begin{figure}[h!]
	\centering
	\subfloat[\textit{interFoam} with kEpsilon \label{fig:interFoam}]{\includegraphics[width=0.49\linewidth, clip = true, trim = 400 100 400 100]
		{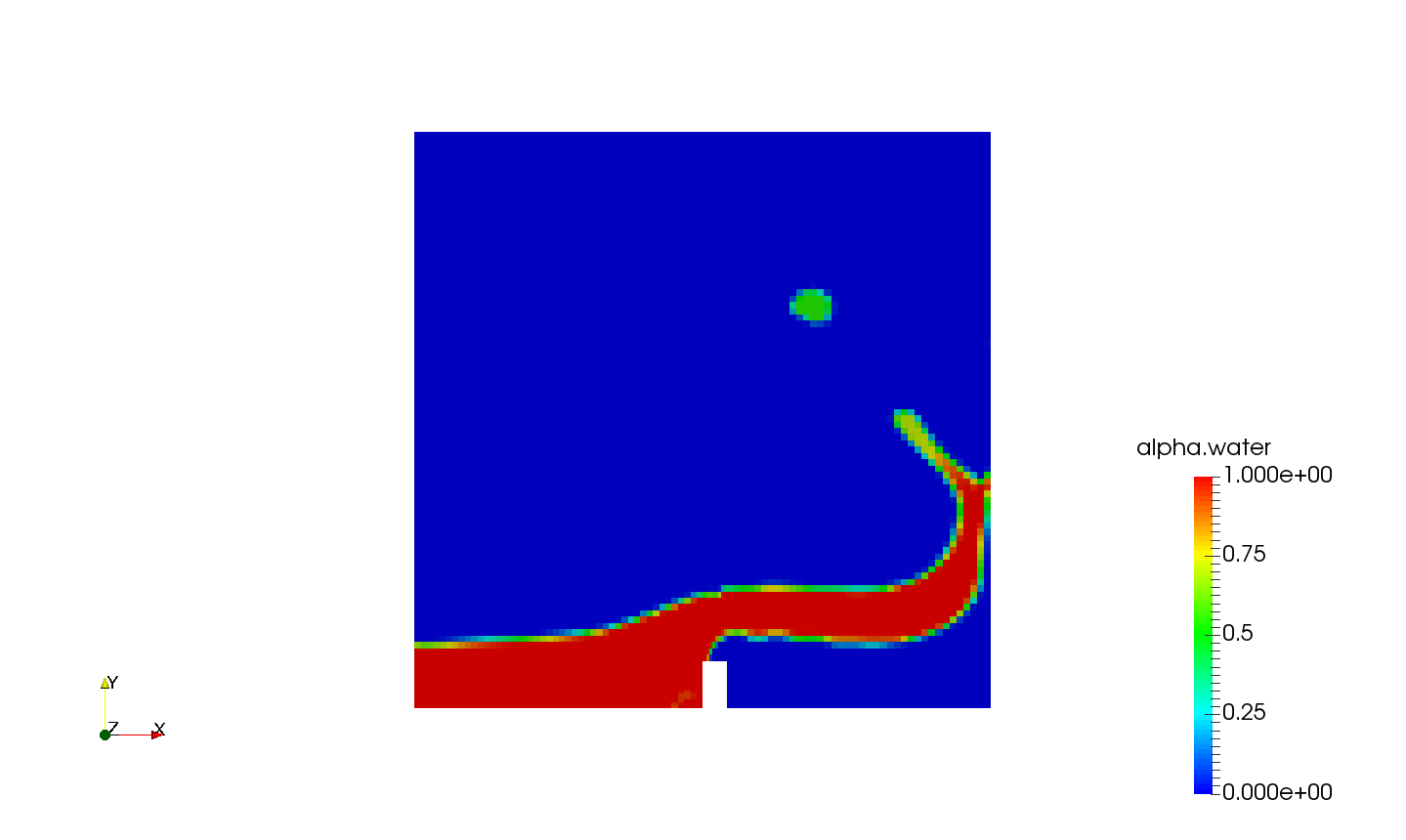}}
	\subfloat[\textit{interFoam} with VOFKEpsilonIm \label{fig:implicit}]{\includegraphics[width=0.49\linewidth, clip = true, trim = 400 100 400 100]
		{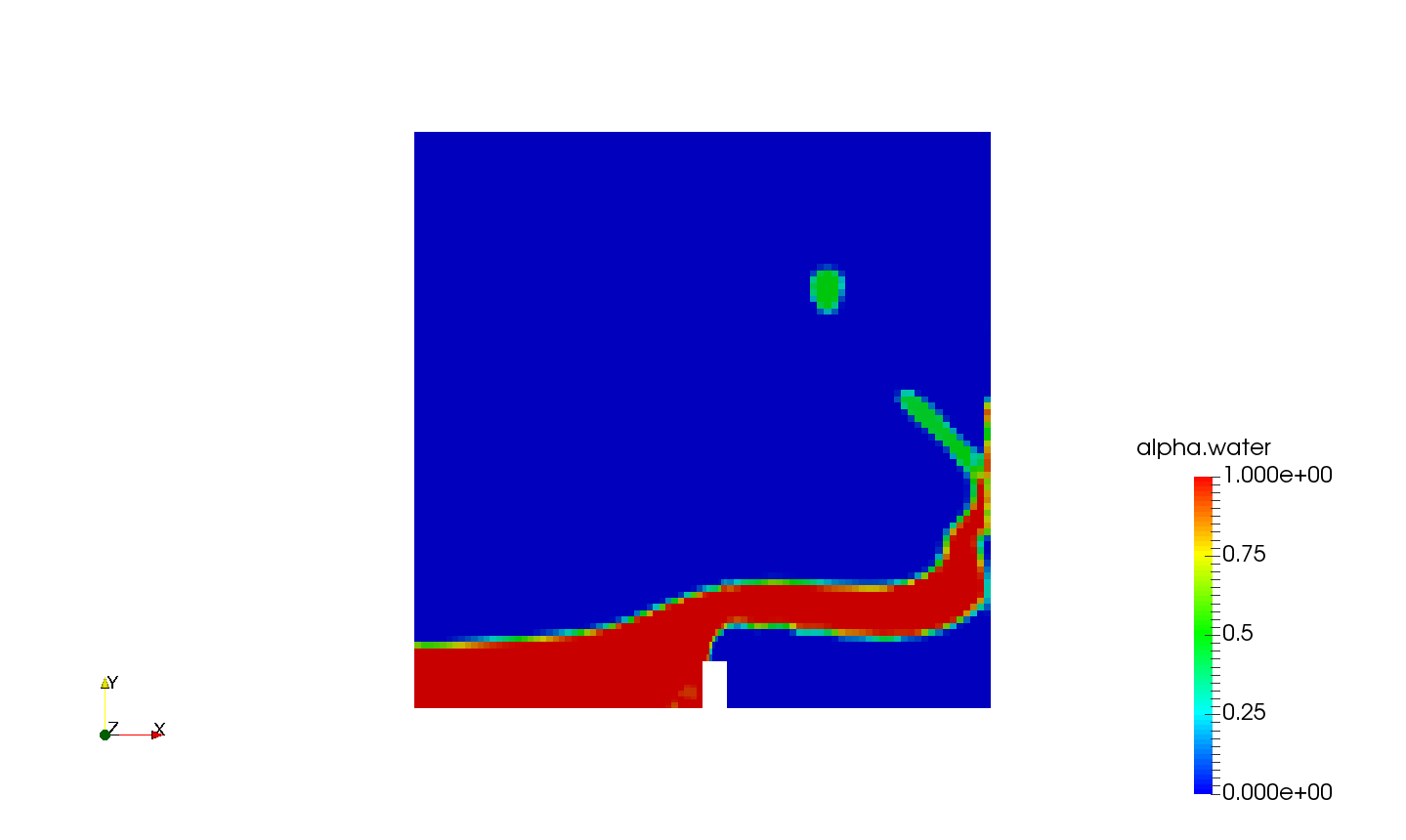}}
	
	\subfloat[\textit{interFoam} with VOFKEpsilonFull \label{fig:full}]{\includegraphics[width=0.49\linewidth, clip = true, trim = 400 100 400 100]
		{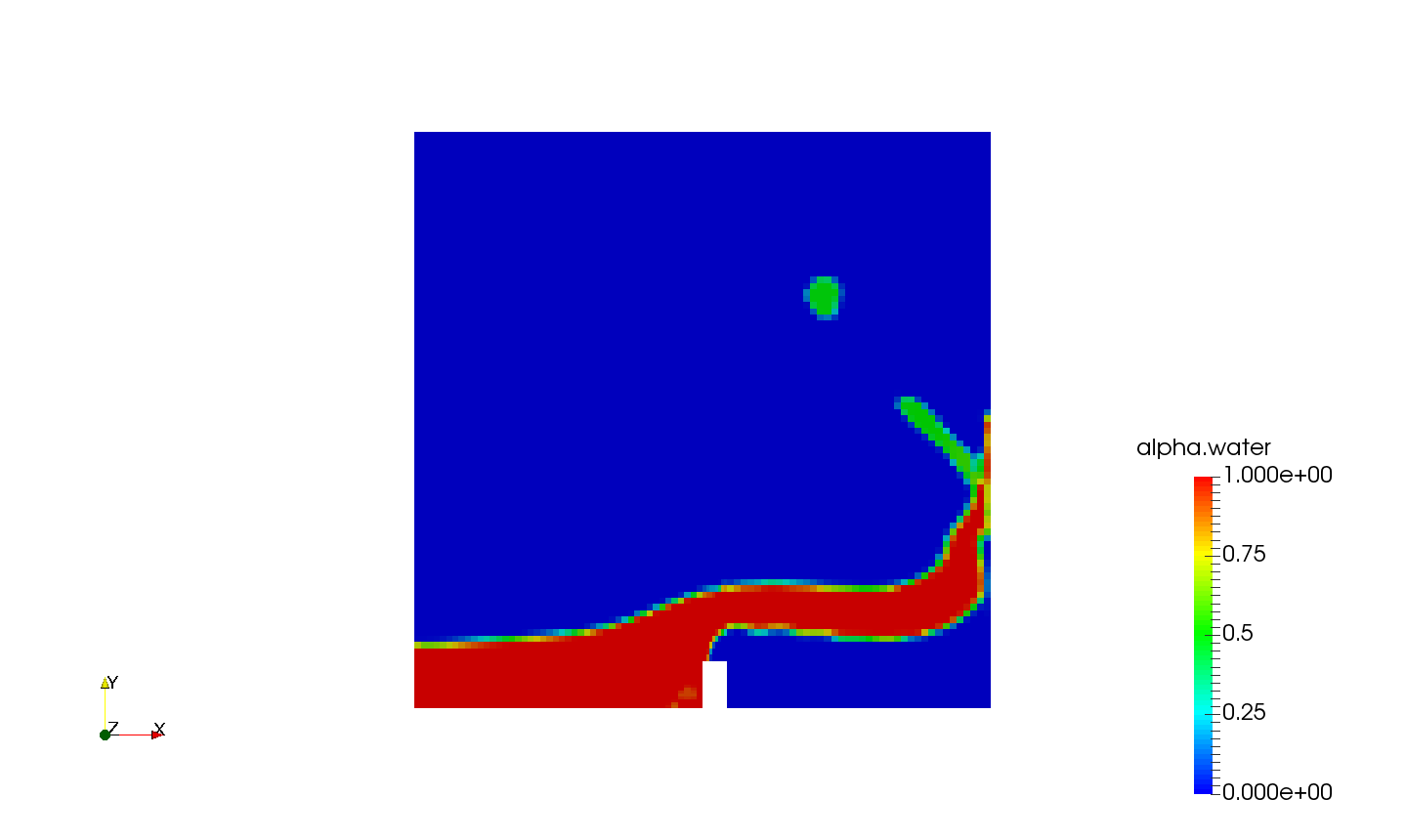}}
	\subfloat[\textit{varRhoInterFoam} with kEpsilon \label{fig:varRho}]{\includegraphics[width=0.49\linewidth, clip = true, trim = 400 100 400 100]
		{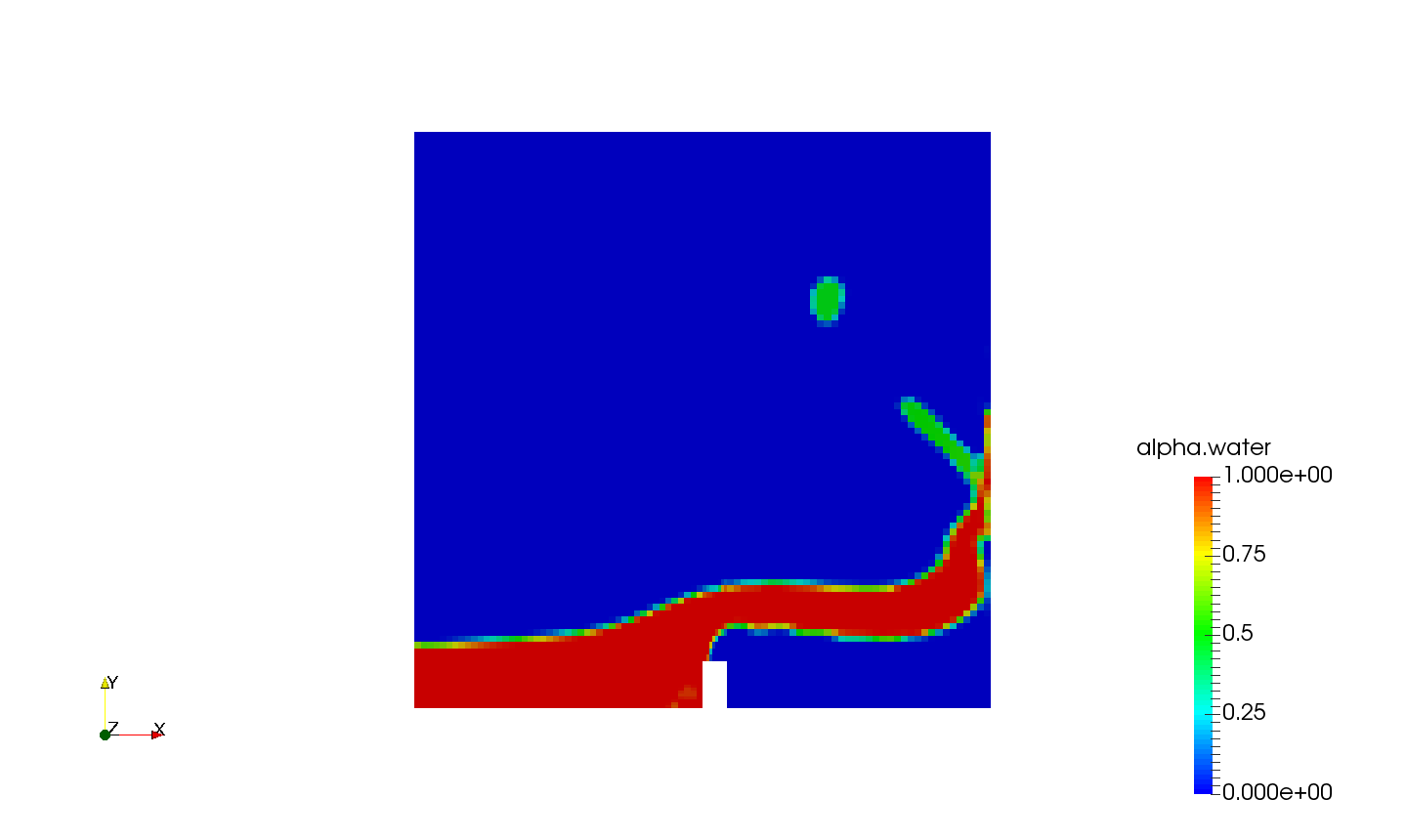}}
	
	\caption{Water-air distribution at the end of simulations, where air and water are denoted by red and blue, respectively.}	
	\label{fig:crossVerification}
\end{figure}

As mentioned in \ref{apdx:alternatives}, we have provided three examples, i.e. VOFKEpsilonEx, VOFKEpsilonIm, and VOFKEpsilonFull, in the source code to show how to create variable-density incompressible turbulence models in a brute-force manner. These models are also useful for the cross-verification test where the classic \textit{damBreak} tutorial is used. The official solver \textit{interFoam} is tested together with these three models and the default kEpsilon model. The newly designed solver \textit{varRhoInterFoam} is only tested in combination with the default kEpsilon model. All the simulations use the same mesh and discretization schemes wherever it is possible. The simulations run for 0.5 s after the dam breaks, and the results are shown in Fig. \ref{fig:crossVerification}. It should be noted that the result for the combination of \textit{interFoam} and VOFKEpsilonEx model is not available due to the crash of the simulation, which is not surprising because of the explicit treatment of the gradient term. Regarding the available results shown in Fig. \ref{fig:crossVerification}, the most obvious differences between the result predicted by the combination of \textit{interFoam} and kEpsilon model ( Fig. \ref{fig:interFoam}) and others are the interaction between the water and the right wall and the orientation of the droplet. The decent similarity among Fig. \ref{fig:implicit}, Fig. \ref{fig:full} and Fig. \ref{fig:varRho} verifies the correctness of code implementation.

\section{More results on performance evaluation}
\label{apdx:result}
\begin{figure}[h!]
	\centering
	\includegraphics[width=0.49\linewidth]
	{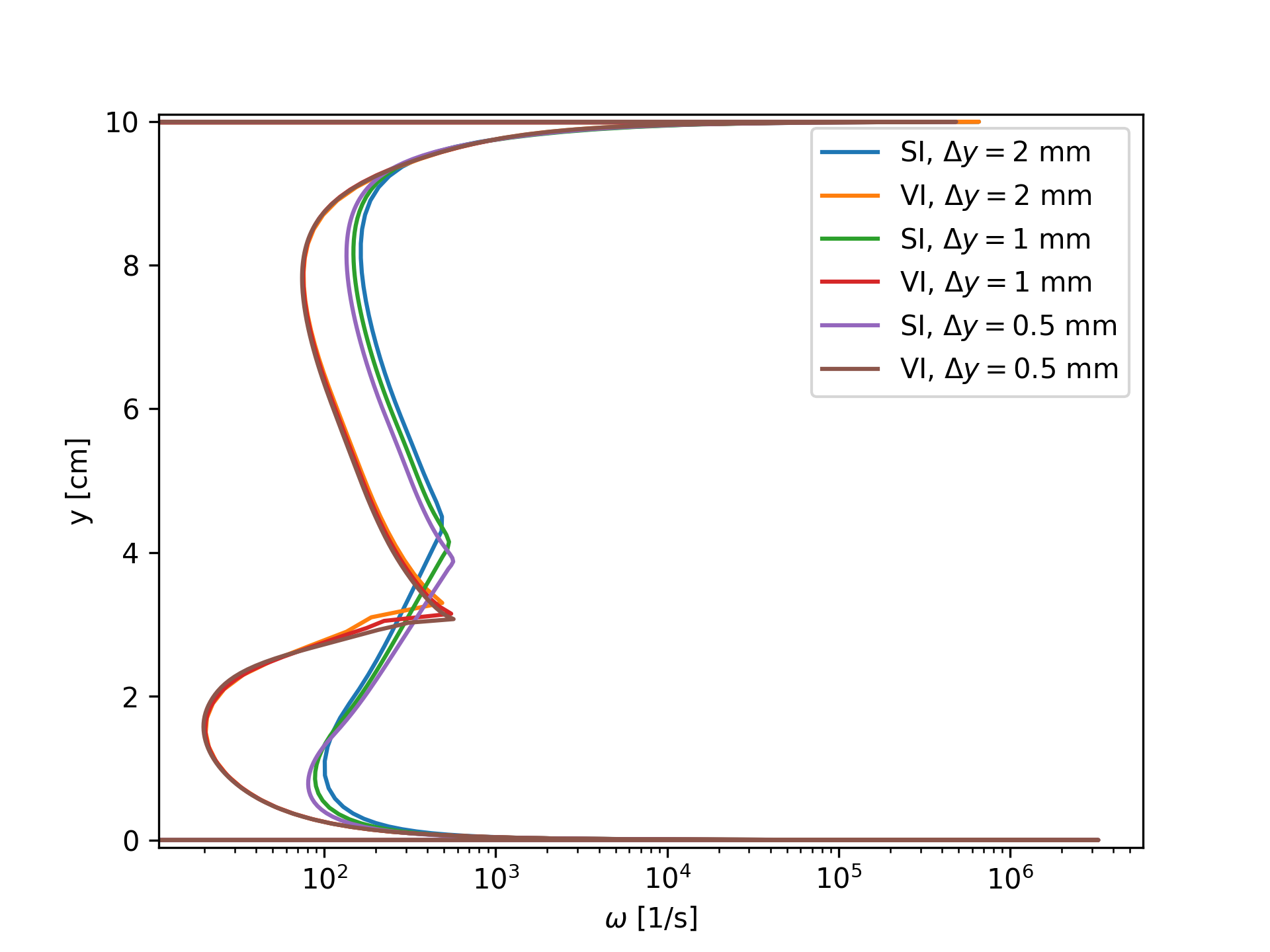}
	\caption{$\omega$ profiles in the vertical direction (run-250).}	
	\label{fig:omega}
\end{figure}

\begin{figure}[h!]
	\centering
	\subfloat [strict incompressilbe \label{fig:strictOmega}]
	{\includegraphics [width=0.49\linewidth] {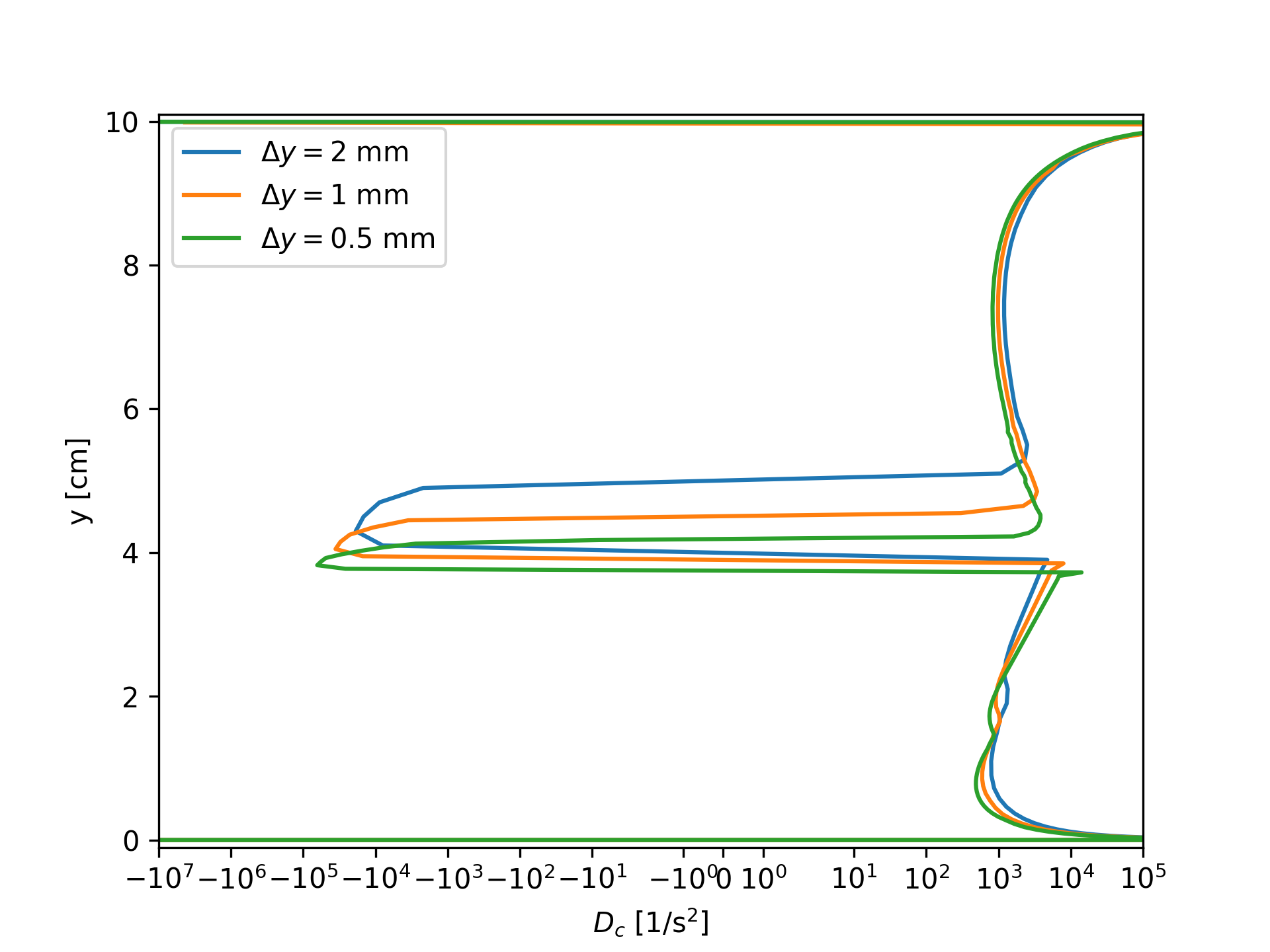}} 
	\subfloat[variable-density incompressible \label{fig:variableOmega}]{\includegraphics [width=0.49\linewidth] {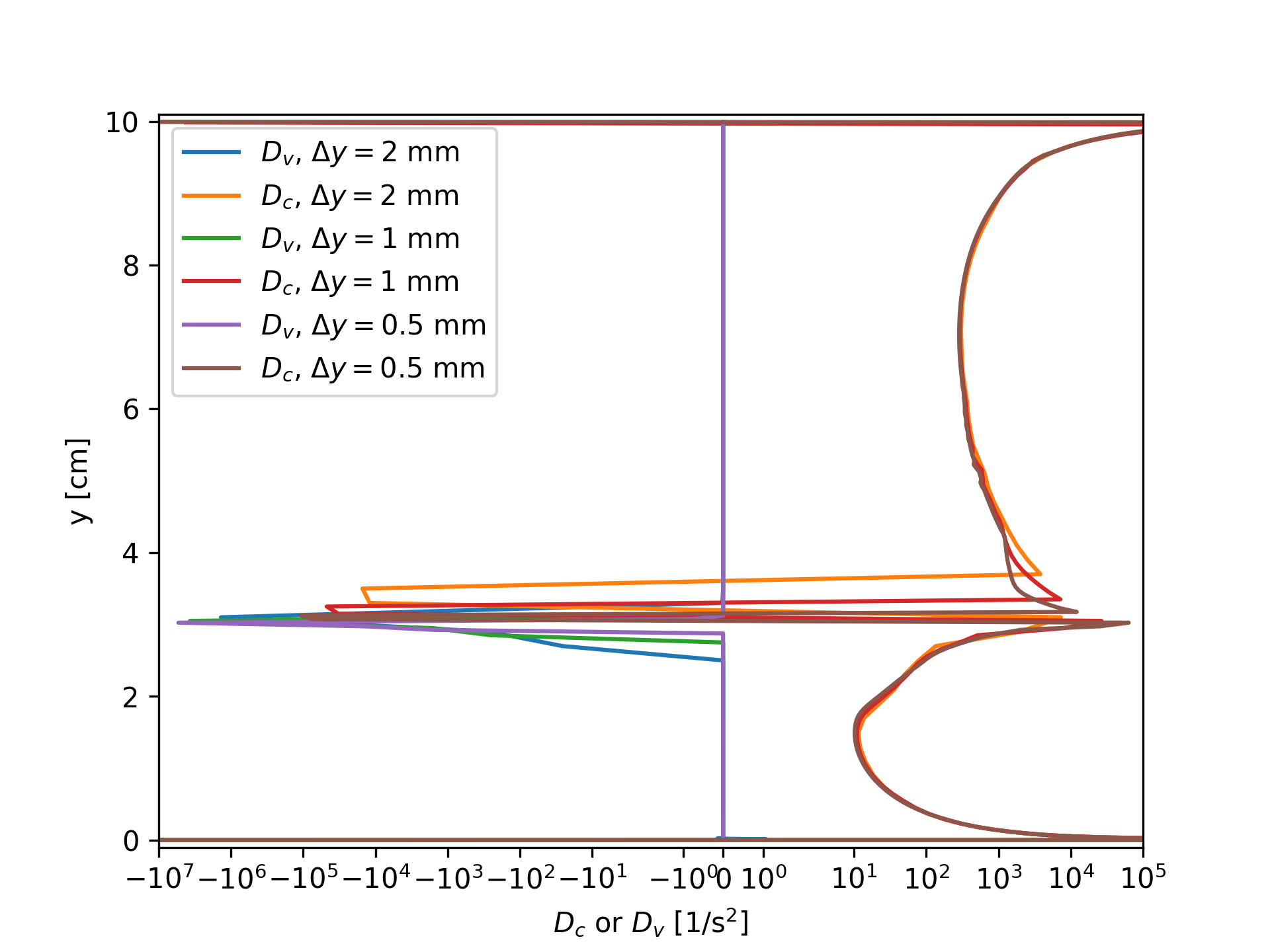}}
	\caption{Comparisons for diffusion terms for $\omega$ (run-250).}
	\label{fig:diffusionOmega}
\end{figure}

Fig. \ref{fig:omega} shows $\omega$ profiles obtained by different models. Even though there are quantitative differences, all the curves have similar shapes. This is quite different from the difference between $k$ profiles shown in Fig. \ref{fig:k250}. The reason could be revealed by inspecting Fig. \ref{fig:strictOmega} and Fig. \ref{fig:variableOmega} where we also split the diffusion term into $D_c$ and $D_v$ parts. For results obtained by the variable-density incompressible model, Fig. \ref{fig:variableOmega} is quite similar to Fig. \ref{fig:variable} around the interface in the sense that the negative $D_v$ terms overweight the positive $D_c$ terms resulting in negative total diffusion terms. For results obtained by the strict incompressible model, unlike diffusion terms for the $k$ equation (Fig. \ref{fig:strict}), diffusion terms for the $\omega$ equation are negative, as shown in Fig. \ref{fig:strictOmega}. Therefore, the total diffusion terms always have a negative sign for both the strict incompressible and the variable-density models. Subsequently, the shape of $\omega$ profiles are not substantially altered in comparison with the shape of $k$ profiles.

\begin{figure}[h!]
	\centering	
	\subfloat[$\nu_t$ \label{fig:nut}]{\includegraphics[width=0.49\linewidth]
		{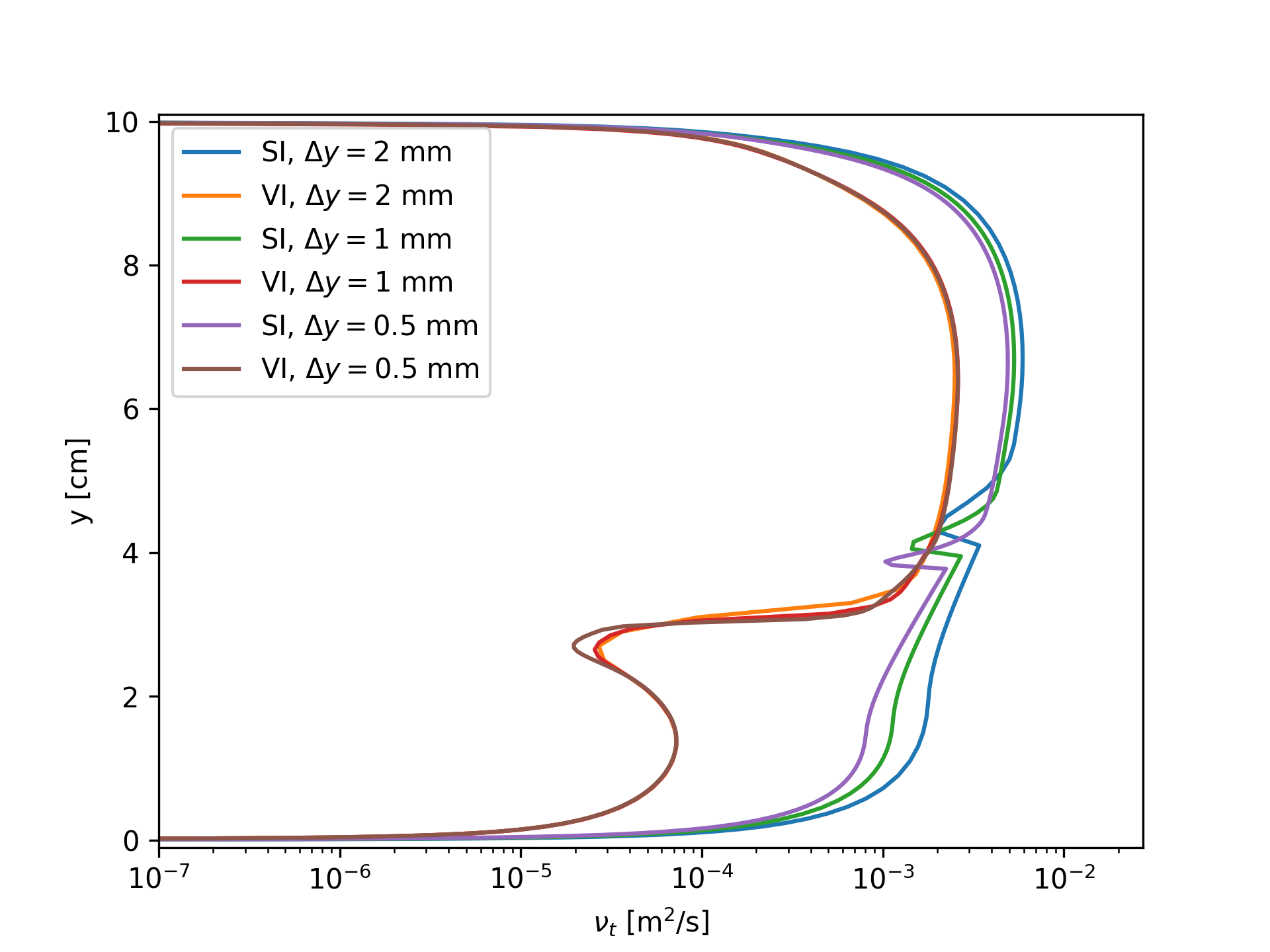}}
	\subfloat[streamwise velocity \label{fig:U}]{\includegraphics[width=0.49\linewidth]
		{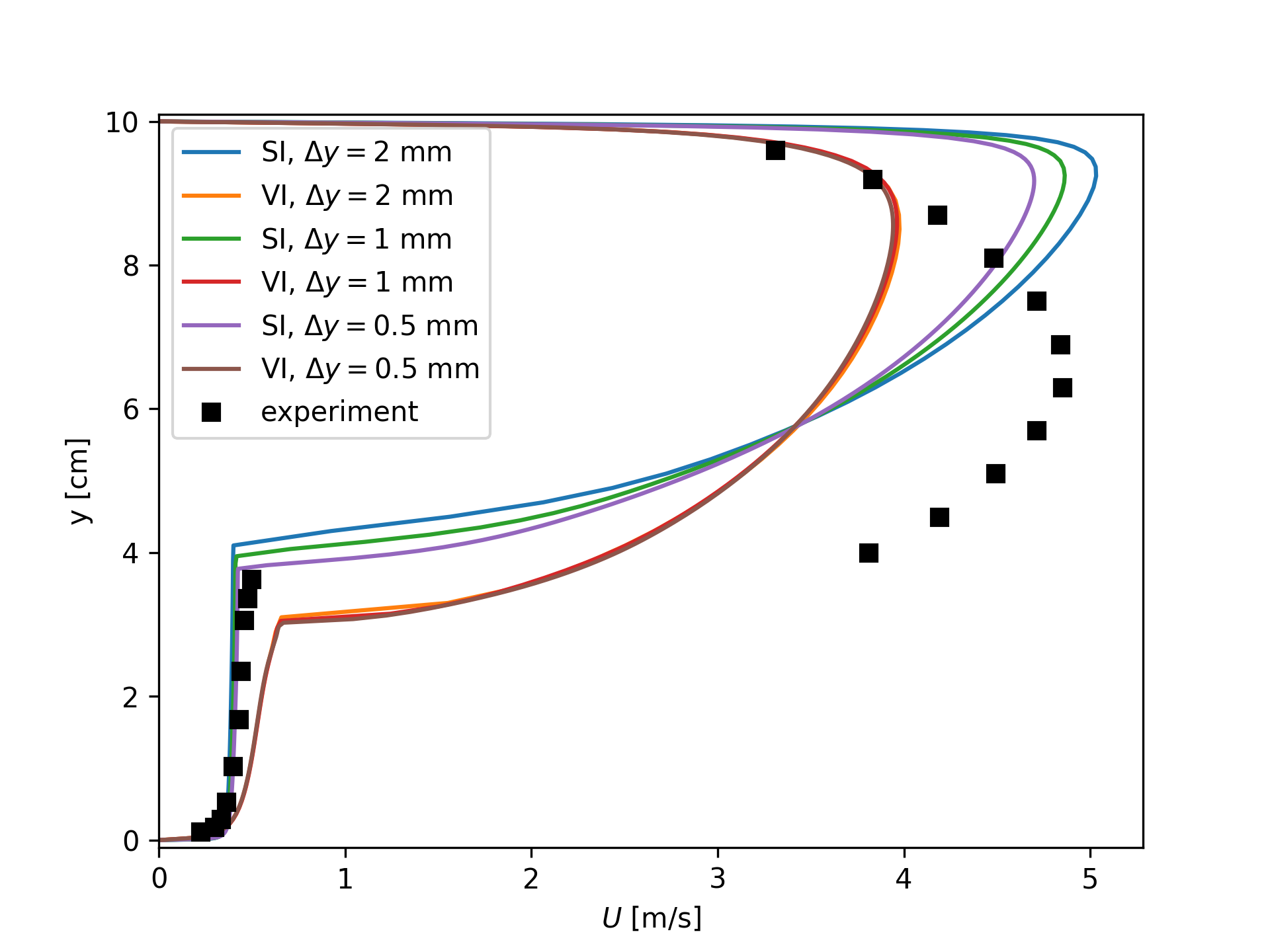}}		
	\caption{$\nu_t$ and $U$ profiles for run-250.}	
	\label{fig:moreResults}
\end{figure}

As mentioned in Section \ref{sect:RANS}, the goal of introducing turbulence models is to calculate the value of $\nu_t$, which is used in the momentum equation. As shown in Fig. \ref{fig:nut}, in comparison with the variable-density one, the strict incompressible calculation predicts higher $\nu_t$ values for both single-phase and interfacial regions. Consequently, velocity profiles are affected as shown in Fig. \ref{fig:U}. It should be noted that, the slope of velocity profiles near the upper wall reflects the pressure gradient of a given result.

\clearpage
	\section*{References}
	
	\bibliography{bib}	
	
\end{document}